\documentclass[10pt,journal,compsoc]{IEEEtran}
\ifCLASSOPTIONcompsoc
  \usepackage[noadjust]{cite}
  
\else
  \usepackage{cite}
\fi
\ifCLASSINFOpdf
   \usepackage[pdftex]{graphicx}
   \graphicspath{{figures/}{pictures/}{images/}{./}}
   \DeclareGraphicsExtensions{.pdf,.jpeg,.png}
\else
\fi
\usepackage{dblfloatfix} 

\usepackage{amsthm} 

\usepackage{caption}
\usepackage{subcaption}

\usepackage{listings}               

\usepackage{flexisym}           

\usepackage[ruled,vlined, linesnumbered]{algorithm2e}  
\SetCommentSty{mycommfont}

\usepackage{enumitem} 

\usepackage{xcolor}

\newcommand{\removed}[1]{\textcolor{red}{\sout{#1}}}

\def \cleanversion{} 
\ifx\cleanversion\undefined
\else

  \renewcommand{\removed}[1]{\iffalse#1\fi}
\fi

\begin{document}
%
\title{Asynchronous and Load-Balanced Union-Find for Distributed and Parallel Scientific Data Visualization and Analysis}
%
%
%
%

\author{Jiayi~Xu, 
        Hanqi~Guo,~\IEEEmembership{Member,~IEEE,} 
        Han-Wei~Shen,~\IEEEmembership{Member,~IEEE,}  
        Mukund~Raj, 
        Xueyun~Wang,
        Xueqiao~Xu, 
        Zhehui~Wang, 
        and~Tom~Peterka,~\IEEEmembership{Member,~IEEE}
\IEEEcompsocitemizethanks{\IEEEcompsocthanksitem Jiayi Xu and Han-Wei Shen are with the Department of Computer Science and Engineering, The Ohio State University, Columbus, OH, 43210, USA.\protect\\
E-mail: \{xu.2205, shen.94\}@osu.edu
\IEEEcompsocthanksitem Hanqi Guo, Mukund Raj, and Tom Peterka are with the Mathematics and Computer Science Division, Argonne National Laboratory, Lemont, IL 60439, USA.\protect\\
E-mail: \{hguo, mraj, tpeterka\}@anl.gov
\IEEEcompsocthanksitem Xueyun Wang is with the School of Physics, Peking University, Beijing, China 100871.\protect\\
E-mail: wxy2015@pku.edu.cn
\IEEEcompsocthanksitem Xueqiao Xu is with the Physical and Life Sciences Directorate, Lawrence Livermore National Laboratory, Livermore, CA 94550, USA.\protect\\
E-mail: xu2@llnl.gov
\IEEEcompsocthanksitem Zhehui Wang is with the Physics Division, Los Alamos National Laboratory, Los Alamos, NM 87545, USA.\protect\\
E-mail: zwang@lanl.gov}
}

\markboth{IEEE Transactions of Visualization and Computer Graphics,~Vol.~X, No.~X, January~2021}%
{Xu \MakeLowercase{\textit{et al.}}: Asynchronous and Load-Balanced Union-Find for Distributed and Parallel Scientific Data Visualization and Analysis}
%



\IEEEtitleabstractindextext{%
\begin{abstract}
We present a novel distributed union-find algorithm that features asynchronous parallelism and k-d tree based load balancing for scalable visualization and analysis of scientific data. Applications of union-find include level set extraction and critical point tracking, but distributed union-find can suffer from high synchronization costs and imbalanced workloads across parallel processes. In this study, we prove that global synchronizations in existing distributed union-find can be eliminated without changing final results, allowing overlapped communications and computations for scalable processing. We also use a k-d tree decomposition to redistribute inputs, in order to improve workload balancing. We benchmark the scalability of our algorithm with up to 1,024 processes using both synthetic and application data. We demonstrate the use of our algorithm in critical point tracking and super-level set extraction with high-speed imaging experiments and fusion plasma simulations, respectively. 
\end{abstract}

}

\maketitle

\IEEEdisplaynontitleabstractindextext

%
\IEEEpeerreviewmaketitle

\IEEEraisesectionheading{\section{Introduction}\label{sec:introduction}}

%
%
%
%


\IEEEPARstart{A}{s} 
the scale of scientific data generated by experiments and simulations grows, it becomes a common practice to use High-Performance Computing (HPC) clusters to analyze and visualize data in parallel. In such distributed and parallel computing environments, data parallelism is a default paradigm; input data are partitioned into data blocks, which are distributed among parallel processes. 
With data-parallelism, intermediate results are produced from individual data blocks before they are merged into the final result. 

This paper focuses on union-find, which is widely used in many scientific visualization algorithms~\cite{carr2003computing, pascucci2004parallel, morozov2013distributed, morozov2014distributed, harrison2015, mcclure2016asynchronous, friederici2019distributed, nigmetov2019local} and plays a key role in merging disjoint sets. 
For example, in the extraction of super-level sets~\cite{friederici2019distributed}, regions with scalar values larger than a given threshold are extracted. Specifically, in 3D volumetric data, union-find merges neighboring voxels that are greater than the given threshold into connected components. 
In critical point tracking~\cite{tricoche2002topology, garth2004tracking}, union-find connects critical points detected in a spacetime gird, which allows effective tracking of time-dependent phenomena. 

In the parallelization of the abovementioned visualization algorithms, we identify two scalability bottlenecks in distributed union-find: (1) high synchronization costs and (2) imbalanced workloads. 
First, because each disjoint set is usually distributed in a subgroup of processes, the use of global synchronizations may block the rest of the processes, causing busy waits in program execution. 
Up to date, distributed union-find~\cite{cybenko1988practical, manne2009scalable, harrison2015, iverson2015evaluation, friederici2019distributed} has been implemented with the bulk-synchronous parallel programming model~\cite{valiant1990bridging}, which manages parallel processes to alternate local computations and global synchronizations iteratively until distributed algorithms converge. In the context of distributed union-find, local computations consist of performing set operations on local data, and global synchronizations are used for merging and updating disjoint sets across processes. 
Second, imbalanced workloads between parallel processes lead to additional busy waits. 
The workload imbalance is caused by processes that possess imbalanced disjoint-set elements. For example, in super-level set extraction, data blocks in some processes may find more voxels above the given threshold than others. Likewise, in critical point tracking, critical points may be non-uniformly distributed in a domain. In this work, we present novel solutions to reduce synchronization costs and balance processes' workloads for distributed union-find, as described below.

First, we eliminate the use of global synchronizations for distributed union-find based on the fact that set unitings are order-independent. 
We prove that it is possible to eliminate global synchronizations by overlapping synchronizations with local computations and guarantee algorithm convergence and final-result correctness. 
It enables the use of asynchronous point-to-point communications to merge and update disjoint sets across processes in practice, which can be fully overlapped with local computations. 

Second, we balance the disjoint-set elements in each process for workload balancing in distributed union-find because the time complexity of local computation is proportional to the number of set elements~\cite{galil1991data}.  
In our implementation, a k-d tree space decomposition is used to redistribute the set elements evenly among processes because k-d trees can be used to effectively balance spatial data, which is the common type for scientific datasets.

We demonstrate the scalability of our distributed union-find algorithm by measuring the performance with up to $1,024$ processes for both critical point tracking and super-level set extraction. 
Benchmark datasets include 3D spatial synthetic data and 2D time-varying application data output by high-speed imaging and fusion plasma simulations. 
We show that our algorithm achieves shorter execution time and better scalability than the existing distributed union-find methods in scientific datasets. In summary, the main contributions of this paper are twofold: 
\begin{itemize}
    \item 
    We prove global synchronizations between processes can be eliminated in distributed union-find, and present a method that allows distributed union-find to overlap communications and local computations, which reduces processes' busy-waiting time. 
    \item 
    We redistribute the disjoint-set elements across processes evenly for load balancing of distributed union-find using a k-d tree decomposition scheme, which improves algorithm scalability in scientific applications. 
\end{itemize}

\section{Related Work}
\label{sect:related_works}


We distinguish shared-memory \cite{shiloach1982logn, anderson1991wait} and distributed-memory \cite{cybenko1988practical, manne2009scalable, harrison2015, iverson2015evaluation, friederici2019distributed} parallelization of union-find, and this paper is focused on the distributed-memory settings. 
Shared-memory parallel union-find focuses on computing environments that share the same memory space, such as multi- and many-core processors. 
Distributed parallel union-find is designed to perform computations in independent processes with distributed memory spaces.


\subsection{Distributed Union-Find Algorithms}
\label{sect:background_union_find}
For data visualization and analysis, most distributed union-find algorithms~\cite{cybenko1988practical, manne2009scalable, harrison2015, iverson2015evaluation, friederici2019distributed} are implemented with the bulk synchronous parallelism~\cite{valiant1990bridging}, which has been the orthodox parallel programming model for distributed iterative algorithms since the 1980s. 
In the context of union-find, the bulk synchronous implementations alternate two stages: (1) each process performing serial union-find computations locally and (2) all processes synchronized to merge and update disjoint sets across processes; the two-stage iterations continue until convergence. 
The benefits of using bulk synchronous parallelism include: 
(1) ensuring disjoint sets are consistent among the processes at each iteration, 
(2) ensuring messages are exchanged in a predetermined order without causing deadlocks of communications, and 
(3) offering straightforward termination detection. 
The drawbacks include disallowing overlapped computations and communications.

Differences of existing distributed union-find algorithms mainly vary in two aspects: (1) data distribution and (2) communication patterns.

\textbf{Data distribution: }
There are two ways to distribute data: full replication and data partitioning. 
Full replication refers to duplicating all set elements among participating processes \cite{cybenko1988practical}, while data partitioning refers to subdividing the input data and redistributing the partitioned data among the processes. Harrison et al. \cite{harrison2015} partitioned mesh data and balanced the mesh cells across the processes using a binary space partitioning approach.

\textbf{Communication patterns: }
To merge and update disjoint sets across the processes, existing distributed union-find methods use global synchronizations with three different communication patterns: (1) master/slave, (2) parallel merging, and (3) neighbor exchange. 
First, with the master/slave pattern in~\cite{friederici2019distributed}, after all the processes finish local computations, all inter-process mergings of sets are sent to the master process to resolve. Then, the master process broadcasts the sets after the mergings to all processes. 
Second, with the parallel merging pattern, Cybenko et al. \cite{cybenko1988practical} built a tree of processes to merge disjoint sets in paired processes every time by sending all disjoint sets of a process to the process's partner. As a result, the root in the tree of processes produces the disjoint sets merged from all processes. 
Third, with the neighbor exchange pattern~\cite{manne2009scalable, harrison2015, iverson2015evaluation}, disjoint sets spanning over adjacent processes are merged and updated at every iteration. The neighbor exchange can be implemented under Message Passing Interface (MPI) standards~\cite{gropp2014using} by using either a collective synchronous operation \texttt{MPI\_Alltoall} involving all processes~\cite{harrison2015} or point-to-point synchronous operations~\cite{manne2009scalable, iverson2015evaluation}.

\subsection{Distributed and Parallel Visualization and Analysis}
\label{sect:back_applications}

Distributed union-find is mostly used for connected component labeling (CCL) in scientific visualization and analysis algorithms, including 
distributed percolation analysis~\cite{friederici2019distributed}, interval volume extraction~\cite{harrison2015}, statistical analysis of connected regions~\cite{mcclure2016asynchronous}, contour tree computation~\cite{pascucci2004parallel, morozov2014distributed}, and merge tree computation~\cite{morozov2013distributed, nigmetov2019local}. 
Although the implementation of CCL~\cite{he2017connected} can be either union-find or breadth-first search~\cite{nguyen2013lightweight, shun2013ligra, galois2020galois, dathathri2018gluon, dathathri2019gluon}, union-find is proved to be more scalable in distributed and parallel settings~\cite{iverson2015evaluation, he2017connected} because union-find organizes elements of connected components using tree structures, explained in Section~\ref{sect:serial_union_find}, which can efficiently synchronize labels of elements in the connected components across processes. 
The rest of this section samples typical scientific visualization and analysis applications with distributed union-find. 






\textbf{Region extraction: }
Visualizing and analyzing spatial regions in scientific data usually offer important insights to scientists, where union-find can play the role of grouping voxels or mesh cells into connected regions. 
In distributed percolation analysis for turbulent flows~\cite{friederici2019distributed}, the percolation function requires quantifying the volume of regions with values higher than a threshold, where the regions are extracted using distributed union-find. 
Interval volume, the volume between two isosurfaces, is important for distributed analysis of supernova simulation, and can be extracted using distributed union-find~\cite{harrison2015}. 
In the distributed statistical analysis of fluid flows~\cite{mcclure2016asynchronous}, connected phase regions are first extracted using the existing distributed union-find~\cite{harrison2015, iverson2015evaluation}; then, each compute node launches asynchronous threads to compute statistics on the extracted regions.

\textbf{Critical point tracking: }
In critical point tracking, union-find is used to connect critical points sharing the same spacetime mesh cells. 
Tricoche et al.~\cite{tricoche2002topology} and Garth et al.~\cite{garth2004tracking} generalized the spatial mesh of the input data into spacetime mesh, with time being the additional dimension. By assuming continuities in the spacetime domain, critical points can be tracked based on spacetime mesh connectivities. 



\textbf{Scalar field topology: }
Distributed union-find has been used in the distributed computation of merge tree~\cite{morozov2013distributed, nigmetov2019local} and contour tree~\cite{pascucci2004parallel, morozov2014distributed}. 
Given scalar values defined on mesh vertices, union-find is used to group vertices on the mesh paths where vertices' values on the paths are increasing.

\section{Preliminaries}
\label{sect:preliminaries}
This section reviews the serial union-find and the bulk synchronous parallelism based distributed union-find methods. 

\subsection{Serial Union-Find}
\label{sect:serial_union_find}

In general, the input of union-find algorithms is a graph, $G=<V, E>$, where $V$ is the collection of all elements and $E$ is the collection of all edges between elements. The output consists of disjoint sets between which no connecting edges exist. 
Union-find has two basic operations: \texttt{Union($v,v'$)} and \texttt{Find($v$)}, where \texttt{Union} unites disjoint sets that two edge-connected elements belong to and \texttt{Find} returns the representative element of the set containing the given element. 
A \texttt{Union} operation is performed on every two edge-connected elements to output final disjoint sets. 
To manage disjoint sets efficiently, the internal implementation of union-find uses a tree data structure~\cite{galler1964improved, galil1991data} as described below. 



\textbf{Disjoint-set trees: } 
Each disjoint-set tree corresponds to a disjoint set and has a \textit{root} element, which is the representative element and the identifier of the set. 
Every element has a parent pointer pointing to a \textit{parent} element and is a \textit{child} of that parent. Every root's parent is itself.

\textbf{\texttt{Union} and \texttt{Find} with disjoint-set trees: }
A \texttt{Find($v$)} operation is implemented by following the parent pointers starting from the given element to identify the root. 
A \texttt{Union($v,v'$)} operation is performed on a pair of edge-connected elements using two sub-operations: (1) \textit{set uniting} and (2) \textit{edge passing}. 
If an edge connects two roots, a set uniting is performed to point one root to the other to form a single disjoint-set tree. Otherwise, the edge is passed through parent pointers of the given elements to \texttt{Find} the roots; after the edge is passed to the roots, a set uniting will be invoked so that the sets of the given elements are merged. 
The set uniting can follow either \textit{uniting by rank}~\cite{tarjan1984worst, galil1991data, manne2009scalable, harrison2015} or \textit{uniting by size}~\cite{galil1991data} rule, which unites one set into the other with a higher rank or a greater size.

\textbf{Path compression: } 
Path compression is a process to shorten the paths from elements to roots in disjoint-set trees by pointing elements to either (1) their grandparents \cite{van1977alternative, weide1980datastructures} or (2) their roots \cite{hopcroft1973set}, where the two choices were proved to have the same amortized time complexity in \cite{galil1991data}. Path compression can make \texttt{Find} and the edge passing operations identify elements' roots with fewer iterations, hence, more efficient. 




\subsection{Distributed Bulk-Synchronous Union-Find}

Distributed bulk-synchronous union-find~\cite{cybenko1988practical, manne2009scalable, harrison2015, iverson2015evaluation, nigmetov2019local} iterates over local computations, synchronous communications, and synchronous termination detection. 

\textbf{Local computations: }
Each process is responsible for handling local work, including processing incoming messages, performing \texttt{Union} operations and path compressions for local elements stored in the process, and queuing outgoing messages. 

\textbf{Communications: }
Processes use synchronous communications to perform inter-process (1) set uniting, (2) edge passing, and/or (3) path compression. 
First, the set uniting following uniting by rank or uniting by size rule requires processes' synchronizations to exchange ranks or sizes of sets to be united and avoid creating cycles in resulting disjoint-set trees. 
Second, processes may pass edges of local elements to the processes of elements' parents. The element parents' processes receive the edges and continue passing edges until reaching roots for future set unitings. 
Third, for path compression, processes may issue queries about elements' grandparents to elements' parents. Then, the processes of elements' parents send the information of grandparents back after receiving the queries. 

\textbf{Termination detection: }
A \textit{collective synchronous communication}, such as \texttt{MPI\_Allreduce} used in~\cite{iverson2015evaluation}, is applied to check whether all processes have completed assigned work periodically for iterations to terminate. 
The ``collective'' means all available processes participate in the communication. 
The ``synchronous'' means the communication blocks the participating processes for an agreement of termination and will not return until all the participating processes respond, ensuring all the participating processes agree when it is ready to terminate.

\section{Overview and Design Considerations}
\label{sect:algo_overview}

Each process's input consists of elements, a subset of $ V $, and the input elements' edges. Elements in different processes are not overlapping. Every element has a unique ordinal \textit{identifier (ID)} and numeric coordinates. The output consists of disjoint-set trees distributed among the processes. 

\begin{figure*}[tb]
  \centering 
  \includegraphics[width=\linewidth]{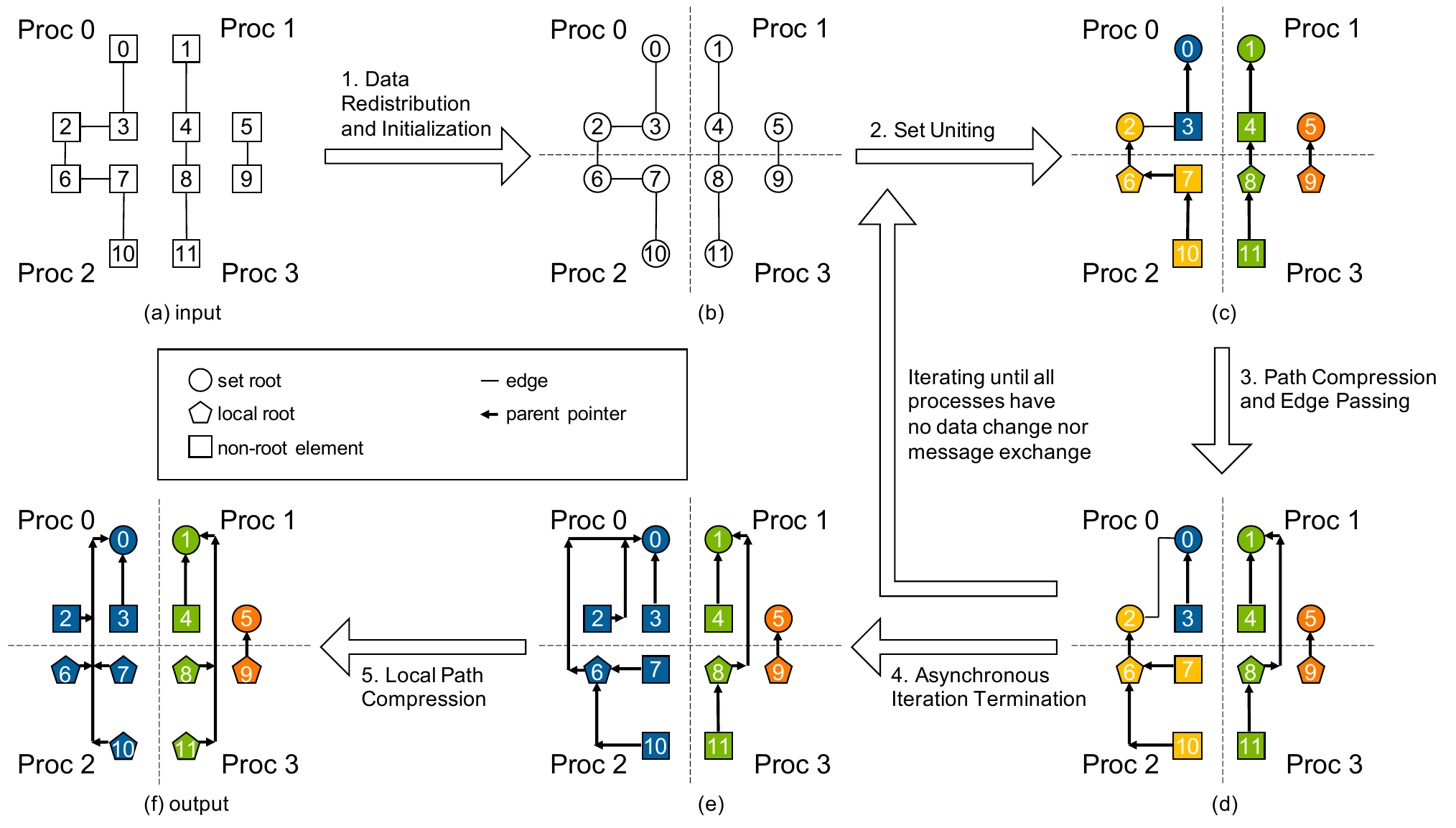}
  \caption{Examples of different operations. 
  (a): An input graph consists of twelve elements. The IDs of the elements are from $0$ to $11$. 
  (b): Elements and edges are redistributed evenly across processes with disjoint sets initialized. 
  (c): Disjoint sets are united in parallel. Colors represent (temporary) disjoint sets. 
  (d): Paths from elements to roots are compressed from (c) to (d), such as the path between $8$ and $1$ and the path between $10$ and $6$. Remaining edges are passed toward set roots; for example, an edge between $3$ and $2$ in (c) is passed to connect $0$ and $2$ in (d). Each process performs the path compression and edge passing independently without blocking other processes. 
  (e): After processes terminate iterations asynchronously, correct disjoint sets are acquired. Certain non-root elements (e.g., $7$, $10$, and $11$) point to local roots rather than set roots. 
  (f): After all non-root elements point to set roots using a local path compression, three disjoint sets are produced. 
  }
  \label{fig:union_find_pipeline}
\end{figure*}

\textbf{Algorithm overview: }
We give examples for involved operations in Fig.~\ref{fig:union_find_pipeline}. 
The algorithm initialization includes redistributing input elements and edges for the load balancing (Section~\ref{sect:load_balancing}) and creating a disjoint set for every element. 
After the initialization, each process iterates over local computations~(Algorithm~\ref{algo:local_computation}), asynchronous communications, and asynchronous termination detection~(Section~\ref{sect:termination_iterations}); each process remains in the iterations as long as there is still remaining local work or incoming new message. 
In the local computations, each process first consumes incoming messages and then repeats set uniting~(Section~\ref{sect:local_union}), path compression~(Section~\ref{sect:async_path_compression}), and edge passing~(Section~\ref{sect:edge_passing}) until local work is complete. The outgoing messages are immediately sent after they have been produced, and the incoming messages promptly drive a new iteration of operations after the incoming messages have been delivered. 
After all processes terminate iterations, we have acquired correct disjoint sets. 
For scientific visualization and analysis, a final local path compression (Section~\ref{sect:async_path_compression}) is performed to label all elements in the same set by the same identifier, i.e., the ID of the set root. 
As not all distributed iterative algorithms can necessarily be designed to overlap communications and local computations, we prove our algorithm's convergence and correctness in~Section~\ref{sect:correctness}.

\textbf{Algorithm design considerations: }
We explain four considerations for the design of distributed asynchronous union-find with the challenges when global synchronizations are eliminated. 

First, we adopt an asynchronous parallelism pipeline with asynchronous communications and asynchronous termination detection to overlap communications and local computations. 
Because asynchronous communications do not block participating processes, processes can continue doing local computations without waiting for communications to complete. 

Second, we use a \textit{uniting by identifier (ID)} rule for the set uniting by uniting a set into the other with a smaller root ID to support overlapping inter-process and local merging of sets. 
Although uniting by rank rule and uniting by size rule are frequently used in the literature, when multiple sets with the same rank or size are united, the two rules may cause cycles in resulting disjoint-set trees and lead to deadlocks if processes are not synchronized~\cite{anderson1991wait}. 
Hence, in shared-memory asynchronous union-find~\cite{anderson1991wait}, when the sets to be united have equal ranks or sizes, Anderson and Woll used the set records in the shared memory to prevent cycles such that one set is united into the other with a greater record index. 
Because distributed-memory processes do not have such shared records of sets, elements' identifiers are used instead in our algorithm.


Third, we create local-tree data structures to reduce communications in contrast to the existing distributed union-find~\cite{cybenko1988practical, manne2009scalable, harrison2015, iverson2015evaluation} without such local trees. 
In our algorithm, only path compression for local-tree roots require communications with the set roots' processes, and the path compression for non-root elements does not involve communications. 
For example, in Fig.~\ref{fig:union_find_pipeline}d, local root $6$ represents the subset consisting of $6$, $7$, and $10$. After $2$ unites with $0$, only the local root $6$ involves communications to point to the new set root $0$ for path compression using our algorithm; however, without local trees, process $0$ may exchange messages with all the three elements ($6$, $7$, and $10$) for path compression, which involves more communications. 


Fourth, we consider boundary cases in path compression to solve deadlocks of communications when the asynchronous termination detection is used. 
After local roots already point to set roots, the local roots may still need to communicate with the set roots for path compression because the set roots may point to other elements after applied union operations. 
For example, in Fig.~\ref{fig:union_find_pipeline}d, local root $6$ may need to keep communicating with set root $2$ for path compression because $2$ may point to another element, i.e., $0$ in this example later. 
A communication deadlock happens when multiple processes with local roots may keep performing such communications with each other even though correct disjoint sets have been built, leading to one of the processes that may always be active so that iterations may never finish when using asynchronous termination detection. 
We present a solution in~Section~\ref{sect:async_path_compression}. 


\section{Distributed Asynchronous Union-Find}
\label{sect:asynchronous_union_find}

We detail our distributed asynchronous union-find. 

\subsection{Distributed Disjoint-Set Trees}

We extend serial disjoint-set trees to create additional data structures for our distributed union-find algorithm. 
    
    \textit{Local disjoint-set trees} are local subtrees formed by local elements of processes, where the \textit{local elements} of a process are elements stored in the process. 
    \textit{Local roots} represent the roots of such local subtrees. Each process can identify local roots by examining local elements that have non-local parents. 
    For example, in Fig.~\ref{fig:union_find_pipeline}c, $6$, $7$, and $10$ form a subtree in process $2$, and $6$ is the local root of the subtree. 
    
    
    A \textit{distributed disjoint-set tree} is formed by all elements of the corresponding disjoint set and may consist of multiple local trees. 
    The roots of the distributed disjoint-set trees are called \textit{disjoint-set tree roots}, or \textit{set root} in short hereafter. 
    Each process can identify the set roots by checking local elements that point to themselves. 
    
    \textit{Non-root elements} represent the elements that are neither local roots nor set roots. 



\subsection{Distributed Union Operations}
\label{sect:local_union}
\label{sect:edge_passing}

Similar to the serial union-find, a distributed \texttt{Union} operation is performed on a pair of edge-connected elements with two sub-operations: (1) distributed set uniting and (2) distributed edge passing. 
Distributed edges are passed to set roots following paths in distributed disjoint-set trees so as to be used for set unitings. 







\textbf{Distributed set uniting: }
The set uniting is based on a \textit{uniting by ID} rule, where two disjoint sets are united by setting the parent of the root of one set to an element with a smaller ID in the other set. 
For example, from Fig.~\ref{fig:union_find_pipeline} (b) to (c), element $6$ sets its parent pointer to element $2$; however, $2$ will not point to $3$ because $3$ is bigger than $2$, and hence, $2$ becomes a (temporary) set root. 
As a result, a parent has a strictly smaller ID than its children in disjoint-set trees, and each set root is the smallest element in a disjoint set, which ensures no cycles. 

The set uniting does not produce outgoing messages. In our implementation, each process points every stored set root to the smallest neighbor element if the process stores or receives any edge connecting the set root with a smaller element. Examples are illustrated in Fig.~\ref{fig:union_find_pipeline}c. 

To make elements be aware of smaller neighbor elements for set unitings, we store each edge in the same process of its larger endpoint after the data redistribution (Section~\ref{sect:load_balancing}) and during the following distributed edge passings. 
For example, in Fig.~\ref{fig:union_find_pipeline}b, the edge between element $5$ and element $9$ is stored in process $3$ so that $9$ is aware of $5$ and can point to $5$ for the set uniting. 





\textbf{Distributed edge passing: }
Edge-passing transports edges of elements to set roots so that the roots are aware of the edges connecting to other disjoint sets for following set unitings. 

Edges are passed through endpoints' parent pointers as follows repeatedly. 
We replace an edge's larger endpoint with the endpoint's parent. 
For example, an edge between element $3$ and element $2$ in Fig.~\ref{fig:union_find_pipeline}c is updated so that connecting $0$ and $2$ in Fig.~\ref{fig:union_find_pipeline}d after the replacement. 
We deprecate the edge if its current endpoints have the same ID; otherwise, the edge is sent to and stored in the process of its current larger endpoint for a set uniting or additional passings. 





\subsection{Distributed Path Compression} 
\label{sect:async_path_compression}

Path compression makes edge passing more efficient by shortening the paths between elements and those connected in the disjoint-set trees with lower IDs until each element's parent pointer points to either a set root or a local root. 
We detail the path compression algorithm for different types of elements as below. 

%


\textbf{Non-root elements: }
Path compression for non-root elements does not produce outgoing messages either.  
At every iteration, non-root elements modify their parent pointers from their parents to the parents' parents, i.e., their grandparents,  if there exist grandparents in the same processes. For example, non-root element $10$ changes its parent pointer from $7$ to $6$ in Fig.~\ref{fig:union_find_pipeline}d and remains pointing to $6$ until iterations finish.  

After the iterations terminate, each process points all the non-root elements to the current set roots, as illustrated in Fig.~\ref{fig:union_find_pipeline}f;   
as a result, all elements within the same set have a common label: the ID of the set root.

\textbf{Local roots: }
At every iteration, each local root communicates to its parent, which is in a different process, to query its grandparent. The parent who receives the query sends the queried information back to the local root for it to update its parent pointer.
For example, in Fig.~\ref{fig:union_find_pipeline}c, after element $8$ queries its grandparent, element $4$ return element $1$ to element $8$. 
When the local root receives the feedback about its grandparent, the local root updates its parent pointer accordingly. 

A boundary case is when a local root has a set root parent, the set root records the local root's process ID after receiving a grandparent query from the local root and sends feedback indicating not to request the grandparent again, which reduces future communications and avoids possible deadlocks of communications due to the use of asynchronous termination detection. 
If the set root later points to another element, the set root notifies its non-local children, which are guaranteed to be local roots in other processes, for path compression. 
After the notification, the local roots are allowed to send additional grandparent queries if they do not point to set roots. 
For example, in Fig.~\ref{fig:union_find_pipeline}d, set root $2$ records the process ID of local root $6$ after receiving the grandparent query from $6$. 
When set root $2$ later unites with $0$, $2$ notifies $6$ of $0$ for path compression.

\subsection{Asynchronous Nonblocking Communications and Termination Detection}
\label{sect:termination_iterations}

Processes exchange messages using asynchronous communications, which do not block participating processes.

\textbf{Communication protocols: }
Three types of messages are exchanged across processes for edge passing and path compression: 
\begin{description}[font={\normalfont\itshape}] 
    \item [Transferred edge] 
    message is used to transfer the data of an edge. 
    This message contains element IDs and process IDs of the two endpoints of the edge. 
    This message is used in (1) edge redistribution after load balancing and (2) edge passing across processes. 
    \item [Grandparent query] 
    message is used when a local root requests for its grandparent. This message contains the element ID and process ID of the local root. 
    This message is sent from the process of the local root to the process of its parent. 
    \item [Grandparent] 
    message is used to answer a grandparent query issued from a local root. This message either contains the element ID and process ID of the corresponding grandparent or indicates the local root's current parent is a set root and inhibit the local root from sending grandparent queries again. 
    This message is sent from the process of the local root's parent to the process of the local root. 
\end{description}

\textbf{Asynchronous termination detection: }
In our asynchronous union-find algorithm, each process exits from iterations when all processes finish local work and no messages are being exchanged. 
The iteration termination detection is straightforward in bulk-synchronous parallelism but requires careful designs for distributed asynchronous algorithms to ensure each process knows all other processes have finished asynchronously and no messages are being transferred. 
We follow the \texttt{iexchange} module of DIY library~\cite{morozov2016block, morozov2021diy} and the nonblocking termination detection mentioned in \cite{dathathri2019gluon}. 
Each process undergoes four states:  (1) \textit{active}, (2) \textit{idle}, (3) \textit{ready-to-terminate}, and (4) \textit{terminate}, and exchanges their states using asynchronous communications to achieve a consensus for correct termination. We refer to~\cite{dathathri2019gluon} for details about the transitions between the states, which are also summarized in the supplementary appendices.

\begin{algorithm} [t]
\SetAlgoLined
    \tcc{consume incoming messages}
    
    
    \For{each msg in in\_msgs}{
        \If{``transferred edge'' in msg}{
            add\_local\_edge(msg, edges)
        }
        \ElseIf{``grandparent query'' in msg}{
            grandparent $\leftarrow$ retrieve\_grandparent(msg, elements)

            comm.isend(grandparent)
        } 
        \ElseIf{``grandparent'' in msg}{
            \tcc{pass compression of local roots}
        
            point\_to\_grandparent(msg, elements)
        }
    }

    \tcc{perform distributed asynchronous union-find operations}

    \While{have local work} {
        
    
        set\_uniting(elements, edges)
        $~\triangleright${Section~\ref{sect:local_union}}

        grandparent\_queries, grandparents $\leftarrow$ path\_compression(elements)
        $~\triangleright${Section~\ref{sect:async_path_compression}}
        
        comm.isend(grandparent\_queries)
        
        comm.isend(grandparents)
        
        transferred\_edges $\leftarrow$ edge\_passing(elements, edges)
        $~\triangleright${Section~\ref{sect:edge_passing}}
        
        comm.isend(transferred\_edges)
        
        
        
        
    }
    

 \caption{local\_computations(comm, in\_msgs, elements, edges) 
 // The ``comm'' is a communicator, such as \texttt{MPI_COMM_WORLD}. The ``isend'' is a point-to-point asynchronous operation for message sending, such as \texttt{MPI_Isend}.
 }
 \label{algo:local_computation}
\end{algorithm}

\subsection{Convergence and Correctness}
\label{sect:correctness}


We prove our algorithm converges to the correct result with a finite number of iterations in each process. 
Our proof has three assumptions. 
First, input data contain a finite number of elements and edges. 
Second, every message can be delivered, but the delivery order of messages is not guaranteed, for example, when MPI asynchronous communications are used. 
Third, the asynchronous termination detection can inform each process to exit iterations when all processes finish local work and no message is being transferred.  




\subsubsection{Converging in Finite Iterations}
We explain our algorithm converges within a finite number of total iterations of processes, where the convergence is achieved when (1) all edges are consumed and (2) disjoint-set trees have no further changes. 


\textbf{(1) All input edges are consumed within finite total iterations. }
\begin{proof}
At each iteration of a process, if a set root has edges connecting with smaller elements, at least one edge is consumed for set uniting; all other edges are passed to a new set root. 
When passing edges, any edge is either (1) deprecated because the endpoints already belong to the same set or (2) passed to a new set root within finite such passes because each disjoint set tree has no cycles, guaranteed by the uniting by ID rule, and contains finite elements. Each pass is either a local operation completed in one iteration or using a message delivered within finite iterations. 
Because set uniting and edge passing keep reducing the number of edges, which is finite, all input edges are consumed within finite total iterations. 
\end{proof}

\textbf{(2) Disjoint-set trees converge to trees with at most two layers within finite iterations, and no messages are exchanged after the convergence of disjoint-set trees.  }
\begin{proof}
In the loop of iterations, disjoint-set trees converge when all non-root elements point to (local/set) roots and all local roots point to set roots.  
Non-root elements point to local/set roots within finite local computations given that, at every iteration, non-root elements of a process will point to their local grandparents for path compression. 
Each local root points to a set root within a finite number of message deliveries, given that the local root points to its grandparent after one grandparent query process. 
Because every message can be delivered within finite iterations, the local root points to a set root within finite iterations. 
Additionally, local roots do not send grandparent queries again after pointing to set roots because the boundary cases in~Section~\ref{sect:async_path_compression} are considered, ensuring no messages are exchanged after disjoint-set trees converge and iterations can properly terminate when asynchronous termination detection is used. 
After the final local path compression, which is illustrated in Fig.~\ref{fig:union_find_pipeline}f and forms one local iteration at each process, all elements point to set roots, leading to trees with at most two layers. 
\end{proof}

\subsubsection{Outputting Correct Disjoint-Set Trees}

The correctness is guaranteed by 
(1) ensuring connected elements in input belong to the same sets in the output and (2) producing unique disjoint set trees; both are independent of the order of set unitings. 

\textbf{(1) Every two edge-connected elements in an input graph belong to the same disjoint set in output. }
\begin{proof}
At the first iteration of each process, if an edge is used for a set uniting, then the two elements connected by the edge are united to the same set; every other edge of the input graph undergoes distributed edge passing. 
Although we cannot guarantee the order of edge passings, we have shown that any edge is consumed within finite iterations. If a passed edge is applied for a union operation, the original two elements of the edge are guaranteed to belong to the same set due to the tree topology. 
If the edge has never been considered for a union operation and is deprecated during the edge passing, the only reason is the original two endpoints of the edge have already belonged to the same set before the edge is passed to set roots. 
Therefore, any two elements connected in an input graph belong to the same set in the output. 
\end{proof}

\textbf{(2) The output consists of ideal disjoint-set trees, where elements, within each set, point to the set's smallest element. }
\begin{proof}
We have shown that, the disjoint-set trees have at most two layers after the convergence, and all elements within a set point to a set root, regardless of the execution order of union operations. 
Because the uniting by ID rule guarantees each disjoint-set tree's root to be the smallest element within each set, our algorithm can produce the ideal disjoint-set trees after convergence. 
\end{proof}


\begin{figure}[htb]
\centering
\includegraphics[width=\linewidth]{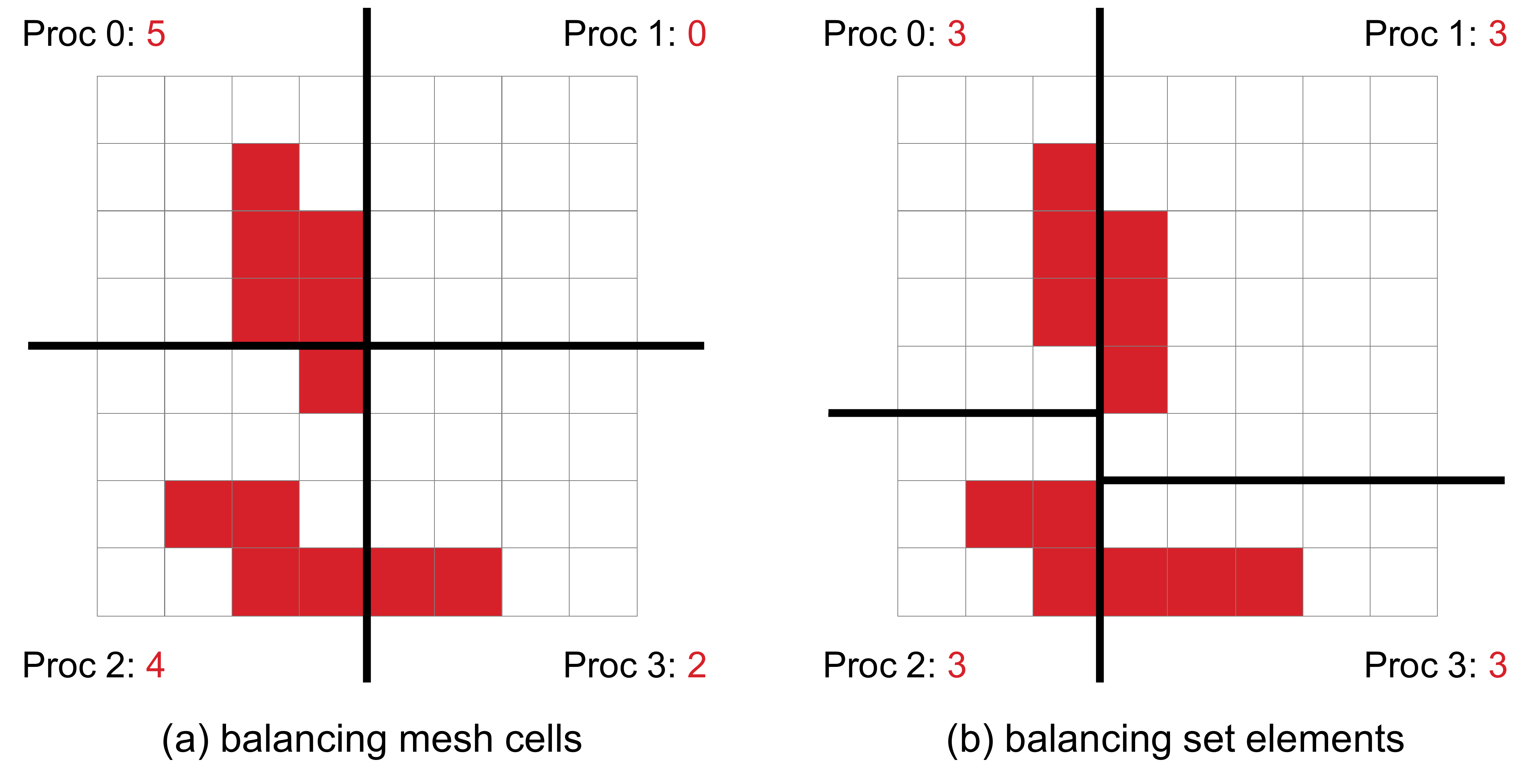}

  \caption{Two possible load balancing schemes for distributed union-find in scientific visualization and analysis. 
  We color set elements by red, and the element counts assigned to processes are indicated in respective corners. 
  (a): Balancing the number of mesh cells in each process using binary space partitioning~\cite{harrison2015}. 
  However, balancing cells may not solve workload imbalance effectively. For example, Process $1$ has zero elements, and hence has no work to do. 
  (b): Balancing the number of elements in each process based on a k-d tree decomposition. In this example, each process is assigned three elements and attains a balanced workload. 
  }
  
  \label{fig:load_balancing_diagram}
\end{figure}

\begin{algorithm} [h]
\SetAlgoLined
    

    rank $\leftarrow$ comm.rank()  \tcp*[h]{ID of this process}

    nproc $\leftarrow$ comm.size() \tcp*[h]{process count}
    
    
    ndim $\leftarrow$ domain dimensionality
    
    dim $\leftarrow$ $0$ \tcp*[h]{current splitting dimension}
    
    group $\leftarrow$ $\{0,~1,~...,~$nproc$ - 1\}$ \tcp*[h]{all processes belong to the same group initially}

    \For{$i \in [0,~\log_2($nproc$))$}{
        m $\leftarrow$ select\_median(comm, elements, group, dim)
        
        i\_bit $\leftarrow$ $1 << i$
        
        group $\leftarrow$ $\{x \in $ group$~|~x$ AND i\_bit $=$ rank AND i\_bit$\}$
        
        partner $\leftarrow$ rank XOR i\_bit
        
        elements $\leftarrow$ swap\_elements(comm, partner, elements, m)
        
        
        dim $\leftarrow$ $($dim$ + 1)$ MOD ndim
    }
    edges $\leftarrow$ redistribute\_edges(comm, elements, edges)
    
    \Return elements, edges
 \caption{balancing\_elements(comm, elements, edges)}
 \label{algo:k-d_tree}
\end{algorithm}
\vspace{-1.5em}

\section{Load Balancing Using K-D Tree based Element Redistribution}
\label{sect:load_balancing}

We redistribute elements to balance the number of elements in each process for workload balancing in distributed union-find, as illustrated in Fig.~\ref{fig:load_balancing_diagram}b. 
We follow the k-d tree space decomposition method proposed by Morozov and Peterka~\cite{morozov2016efficient} for distributed computing environments. Modified pseudocode is presented in Algorithm~\ref{algo:k-d_tree} and described below to support the load balancing in distributed union-find. 


Given distributed elements with numeric coordinates, the decomposition approach partitions the domain so that each process is assigned a partition that contains a similar number of elements. 
Initially, all processes belong to a single group, which represents the whole domain. 
The domain is then decomposed based on medians of the coordinates of elements. After a median is selected, the group of processes is split into two subgroups: the first subgroup contains all processes whose IDs' least significant bit is $0$, and the second subgroup contains all processes whose IDs' least significant bit is $1$. 
Elements are then exchanged between the two subgroups so that the first subgroup corresponds to the lower half of the domain and the second subgroup corresponds to the upper half of the domain. Each process in the first subgroup sends local elements that lie above the selected median to a partner process in the second subgroup; the partner process shares the same bits in the ID as the first group process except that the least significant bit is $1$. Also, the partner process sends the first group process the elements below the selected median. 
In the following iterations, the domain decomposition happens within every subgroup based on the second, third, and so on significant bits of process IDs until each subgroup contains one process. 

After domain decomposition, we redistribute edges such that every edge is stored in its larger endpoint's process.

\section{Algorithm Evaluation}
\label{sect:algorithm_evaluation}

We evaluate distributed union-find algorithms under a scientific feature extraction and tracking framework consisting of four stages: (1) domain partitioning, (2) feature detection, (3) connected component labeling (CCL) using distributed union-find, and (4) finalization, following existing scientific studies in the Feature Tracking Kit (FTK)~\cite{guo2020ftk, guo2019ftk}. 


First, we evenly partition the input spatiotemporal data domain into spacetime blocks, which are then distributed over the participating processes. 
The spacetime blocks are constructed using spacetime meshing \cite{tricoche2002topology, garth2004tracking} to support detecting scientific features with time continuity~\cite{tricoche2002topology} and capturing topological events~\cite{tricoche2002topology, ji2003volume}. 
Each spacetime block includes one layer of ghost cells in both space and time dimensions to associate spatiotemporal features across blocks. 

Second, each process independently detects features and connections between the features.  
For example, in super-level set extraction, we detect spacetime cells with values higher than a specified threshold.  In critical point tracking, we follow Tricoche et al. \cite{tricoche2002topology} to detect critical points on the faces of spacetime cells: we estimate derivatives at mesh grid points using central difference and locate critical points on the faces using inverse interpolation. 
The connections exist between features in adjacent spacetime cells or faces. 


Third, we perform CCL using a distributed union-find algorithm. CCL involves detecting the connectivity between features and labeling each connected component by a unique identifier. 
We regard spatiotemporal features as elements in disjoint sets and connections between the features as edges between the elements. 
As a result of the distributed union-find algorithm, features within the same connected component are merged into a single set and labeled by a common identifier. 

Fourth, we finalize feature extraction and tracking by gathering features within the same connected component to the same process for further visualization, analysis, and storage. 

Our study below focuses on measuring the performance of CCL using distributed union-find algorithms. 

\textbf{Baseline: }
We implement the distributed union-find (DUF) method of Iverson et al. \cite{iverson2015evaluation} load-balanced by balancing mesh cells~\cite{harrison2015} as the baseline for comparison. 
The study of \cite{iverson2015evaluation} evaluates five state-of-the-art distributed approaches on the CCL task and indicates that, except the breadth-first search based label propagation exhibiting considerably worse results, the other four approaches, including the DUF, have similar strong scaling efficiency in each test data.

Compared with the baseline, the improvement caused by overlapping communications with computations using asynchronous parallelism is demonstrated on synthetic data (Section~\ref{sect:exp_synthetic}). 
The benefit of redistributing elements for load balancing is highlighted in both experimental and simulation datasets due to which have non-uniformly distributed features (Section~\ref{sect:applications}). 

\textbf{Computing platform: }
We run experiments on an HPC cluster, which has $664$ compute nodes. Every compute node has Intel Xeon E5-2695v4 CPUs with $32$ cores and $128$ GB memory. The HPC cluster uses an Intel Omni-Path interconnect network. Message passing is supported by the Intel MPI library. The file storage system of the cluster is IBM General Parallel File System.

\subsection{Benchmark on Synthetic Data}
\label{sect:exp_synthetic}

\begin{figure}[htb]
\centering
\captionsetup[subfigure]{labelformat=empty}

  \begin{subfigure}[b]{0.05\linewidth}
   \includegraphics[width=\linewidth]{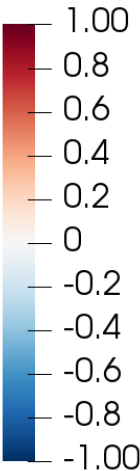}
   \caption{}
 \end{subfigure}
 \begin{subfigure}[b]{0.25\linewidth}
   \includegraphics[width=\linewidth]{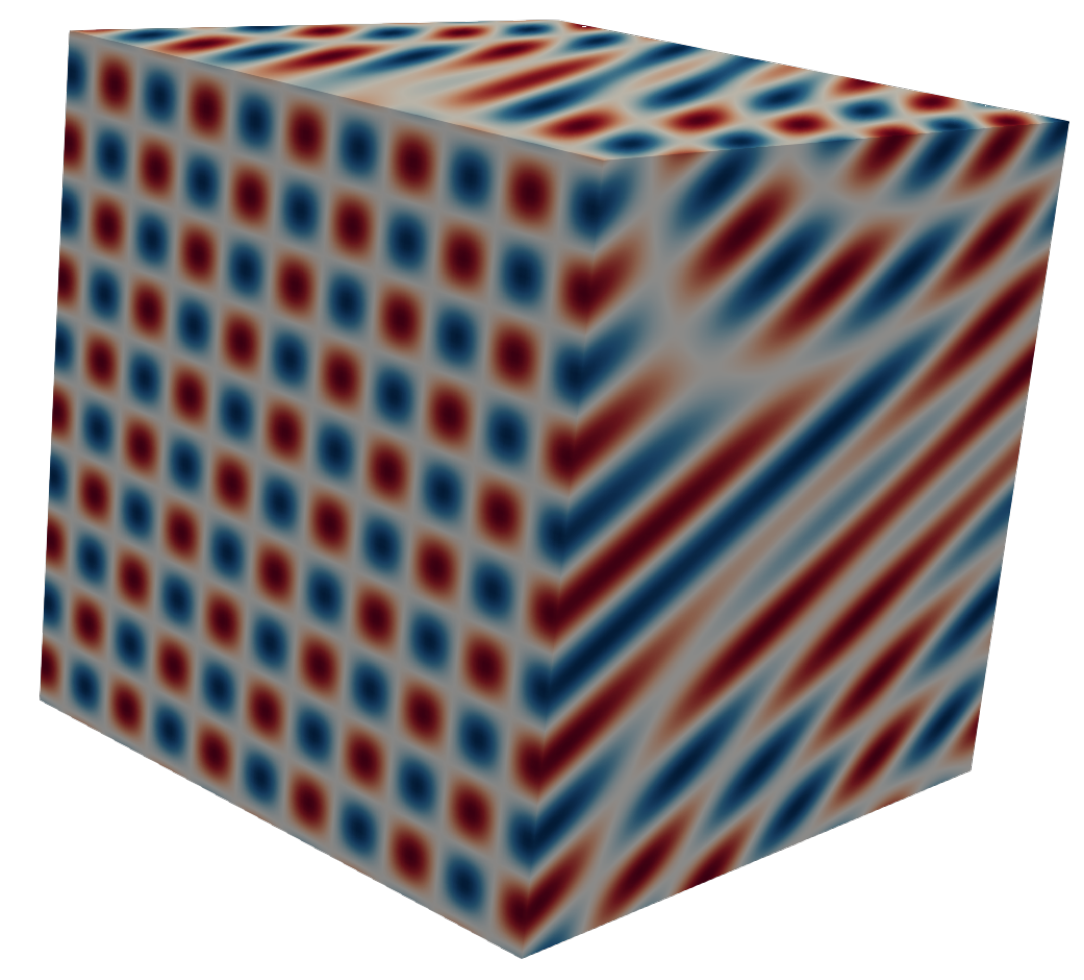}
   \caption{(a)}
 \end{subfigure}
 \begin{subfigure}[b]{0.25\linewidth}
   \includegraphics[width=\linewidth]{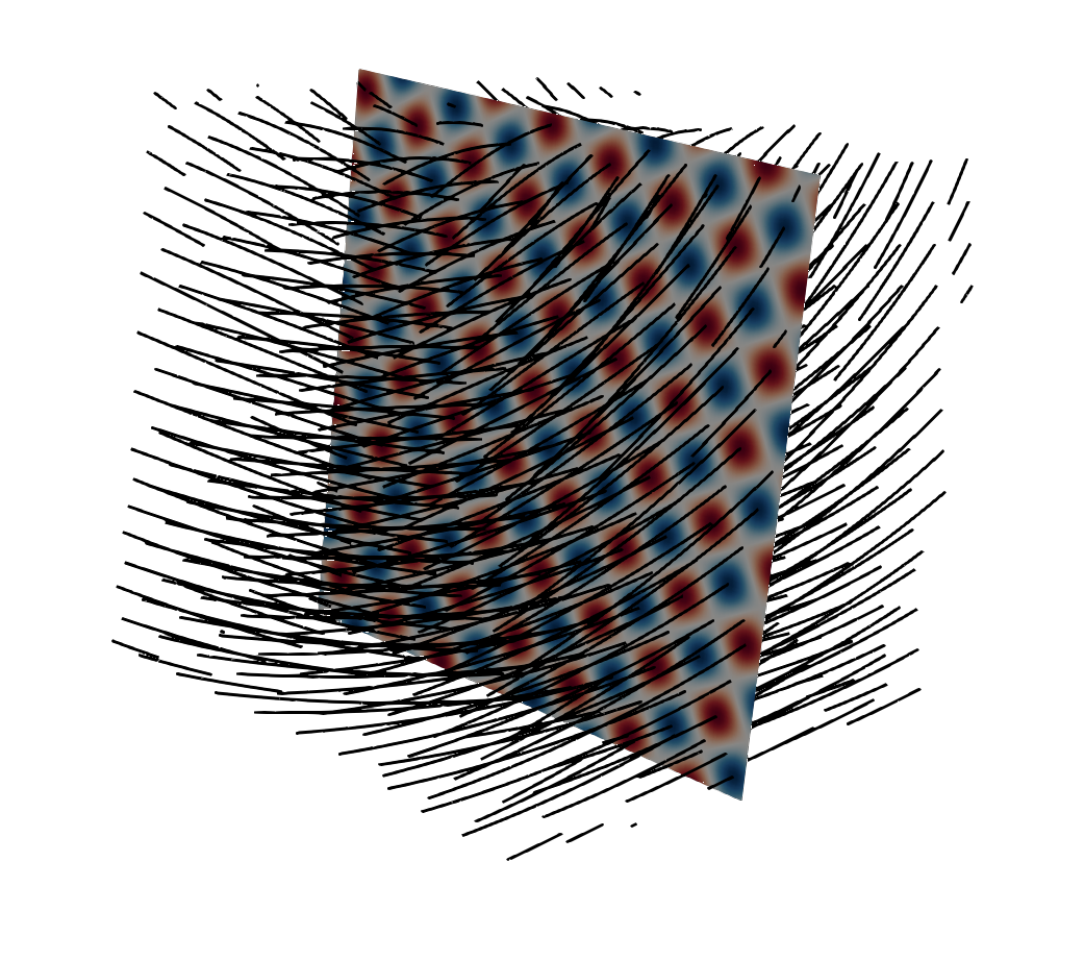}
   \caption{(b)}
 \end{subfigure} 
 \begin{subfigure}[b]{0.25\linewidth}
   \includegraphics[width=\linewidth]{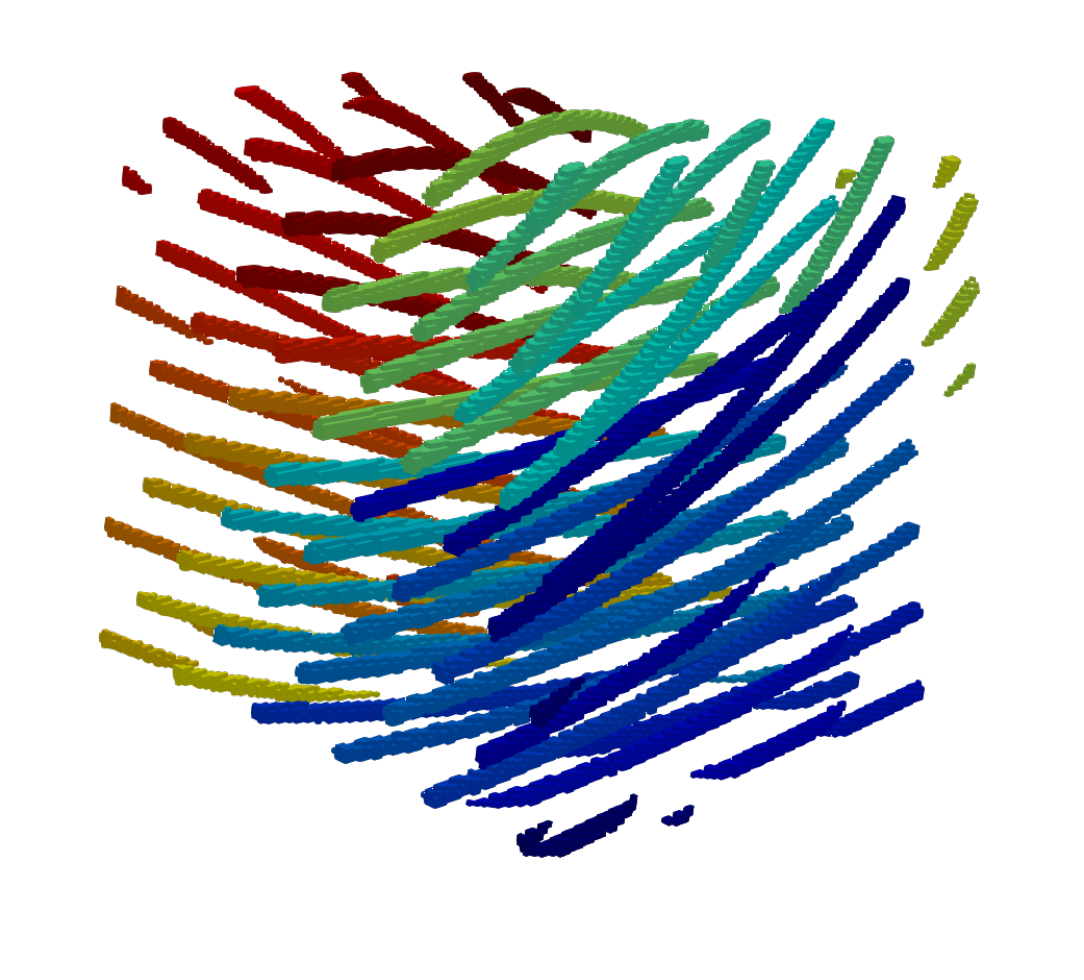}
   \caption{(c)}
 \end{subfigure} 
 \begin{subfigure}[b]{0.05\linewidth}
   \includegraphics[width=\linewidth]{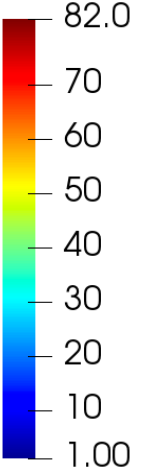}
   \caption{}
 \end{subfigure}

  \caption{Visualizations of a synthetic case. 
  (a): A synthetic volume with scalar values ranging from $-1$ to $1$. We track critical points and extract super-level sets on the same synthetic volume. 
  (b): Trajectories of the critical points represented by black lines. 
  (c): Super-level sets with a threshold of $0.8$. $82$ connected components are labeled, and each is assigned a unique hue. 
  }
  
  \label{fig:synthetic_visualizations}
\end{figure}

We begin by measuring the influence of overlapping communications and computations using asynchronous parallelism with asynchronous communications and asynchronous termination detection for distributed union-find. 
In the experiments, we track and extract two types of features, including (1) critical points (local maxima, local minima, and saddle points) and (2) super-level sets, on synthetic data. Synthesizing data allows us to control data resolution and the number of features 
to support different evaluations. An example of the synthetic data we use is visualized in Fig.~\ref{fig:synthetic_visualizations}.

\begin{figure}[htb]
\centering
 
  \includegraphics[width=\linewidth]{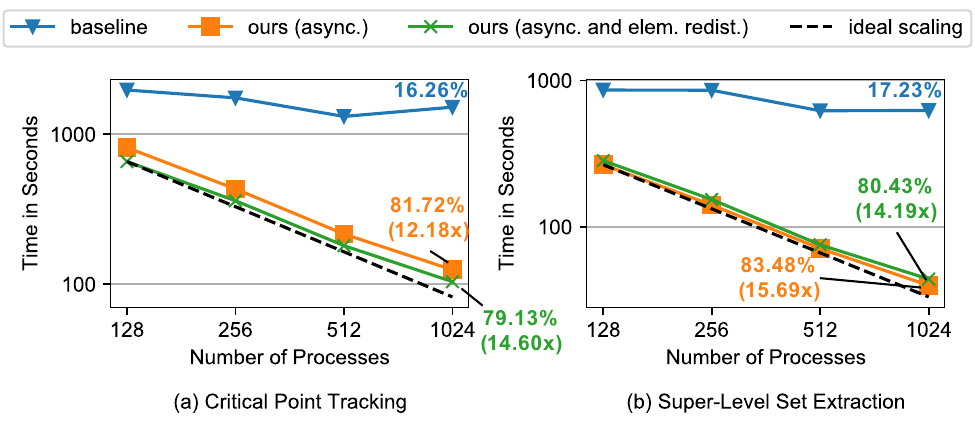}

  \caption{Strong scaling of distributed union-find on $1,024^3$ synthetic data using $128$ to $1,024$ processes for (a) tracking critical points and (b) extracting super-level sets. Both axes are log scales. 
  The baseline is the distributed union-find (DUF) of Iverson et al. \cite{iverson2015evaluation} with balanced mesh cells \cite{harrison2015}. Our methods consist of the distributed asynchronous (async.) union-find without/with the k-d tree based element redistribution (elem. redist.). 
  }
  
  \label{fig:sync_vs_async}
\end{figure}

\begin{figure*}[tb]
\centering

    \includegraphics[width=\linewidth]{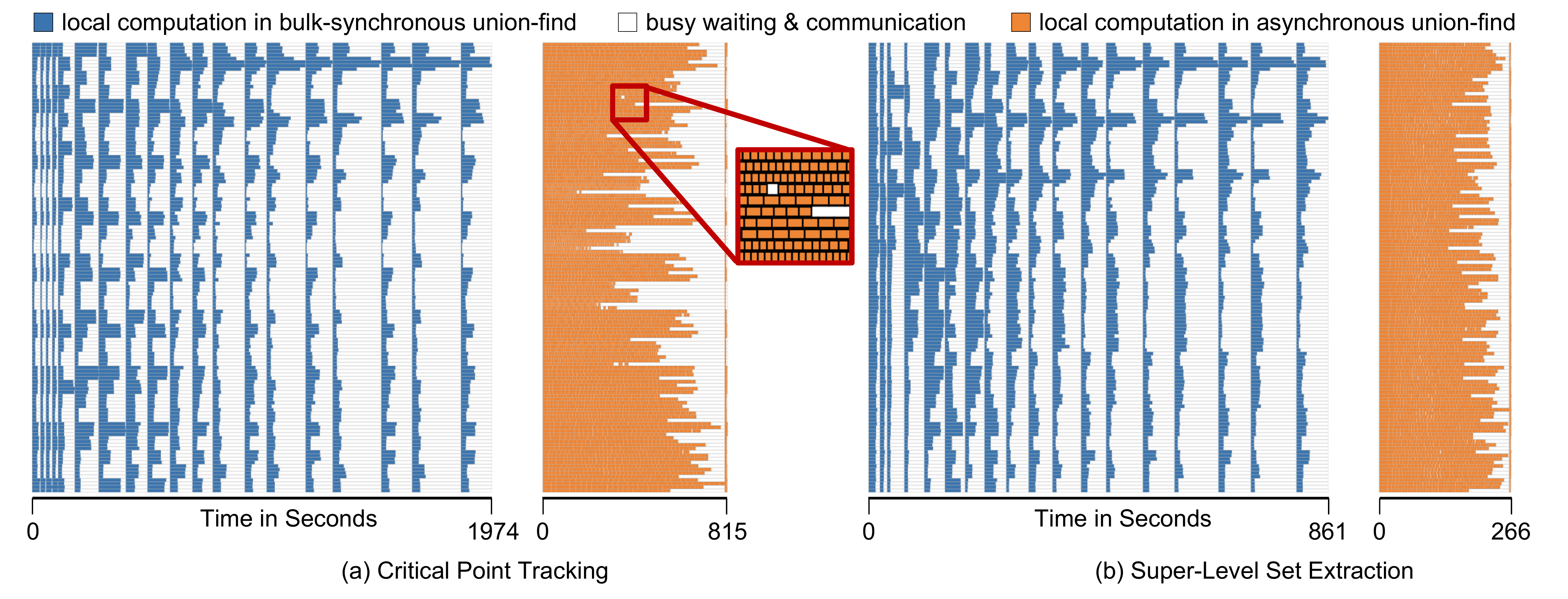}

  \caption{Gantt charts of bulk-synchronous baseline \cite{iverson2015evaluation} and our asynchronous algorithm using $1,024^3$ synthetic data distributed among $128$ processes. 
  The horizontal axis encodes time. 
  Each row corresponds to a process. 
    }
  
  \label{fig:sync_vs_async_breakdown}
\end{figure*}

\subsubsection{Strong Scalability Study}
\label{sect:eval_strong_scaling}

We conduct a strong scaling experiment using synthetic data with a $1,024^3$ resolution, which has $161,338,942$ critical points and $72,097,212$ voxels with values larger than $0.8$. 

Compared with the baseline using the bulk-synchronous parallelism, the use of asynchronous parallelism leads to significant improvement in strong scaling efficiency. 
Strong scaling results using up to $1,024$ processes are shown in Fig.~\ref{fig:sync_vs_async}. 
The baseline attains a strong scaling efficiency of $16.26\%$ in the critical point tracking benchmark and $17.23\%$ in the super-level set extraction benchmark. 
Our algorithm with the asynchronous parallelism attains $81.72\%$ and $83.48\%$ for the two benchmarks, respectively, with a speedup of $12.18$x and $15.69$x over the baseline. 


To investigate why overlapping communications and computations using asynchronous parallelism is better than the distributed bulk-synchronous parallelism in detail, we list processes' computation time at each iteration in Fig.~\ref{fig:sync_vs_async_breakdown}. 
In the distributed bulk-synchronous baseline, a global synchronization is performed at the end of each iteration to merge and update sets across the processes and detect iterations' termination. Due to the use of global synchronizations, processes with less computational work become busy-waiting for processes with more work at each iteration round. 
After overlapping communications and computations using asynchronous communications and asynchronous termination detection, each process performs computations at their own pace without being blocked by other processes, leading to reduced waiting time and improved computational resource usage. After the end of the iterations, our algorithm performs an additional local path compression, however, which occupies only a small portion of the total execution time. 


\begin{figure}[htb]
\centering
 
  \includegraphics[width=\linewidth]{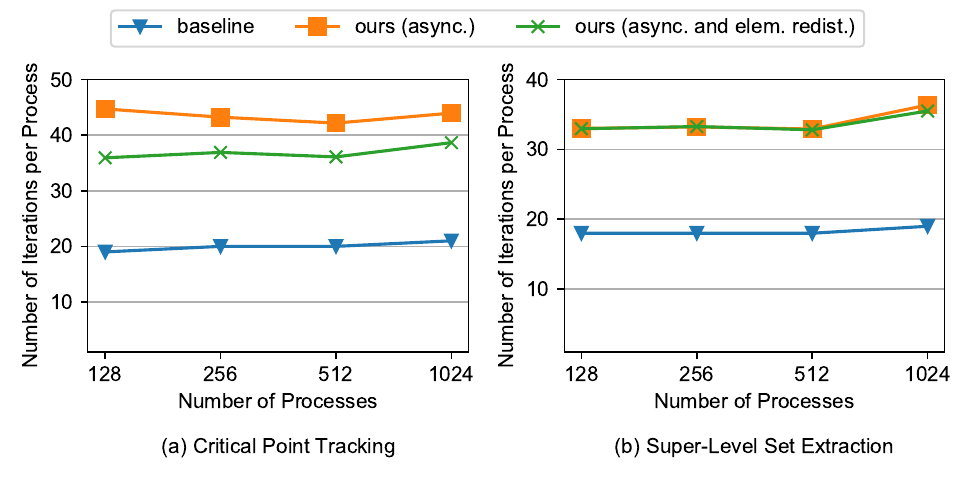}

  \caption{Iteration count per process of distributed union-find on $1,024^3$ synthetic data.
  The horizontal axis is a log scale. 
  }
  
  \label{fig:sync_vs_async_iterations}
\end{figure}

We also evaluate the influence of using asynchronous parallelism on the number of iterations per process. 
Results in Fig.~\ref{fig:sync_vs_async_iterations} illustrate that our distributed asynchronous approach has approximately doubled iterations compared with the bulk-synchronous baseline in both the critical point tracking and super-level set extraction. 
Because asynchronous communications are nonblocking, processes can iterate without waiting for other processes. As long as new messages come, processes can immediately come to the next iteration and handle new work. 
In contrast, the bulk-synchronous parallelism synchronizes processes between iterations. Hence, each process may receive work from all other processes after a synchronization, leading to more work to do during the next iteration and requiring fewer iterations. 
Although using asynchronous parallelism results in more iterations, our algorithm's total execution time is less than the bulk-synchronous baseline, as seen in Fig.~\ref{fig:sync_vs_async}.

\begin{figure}[htb]
\centering


\includegraphics[width=\linewidth]{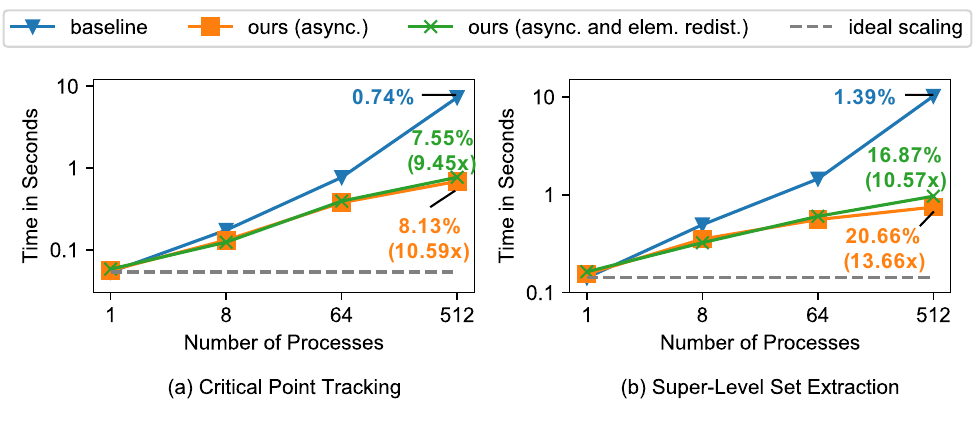}

  \caption{Weak scaling of distributed union-find on synthetic data. 
  We use four combinations of data resolutions and process counts: $32^3$ with $1$ process, $64^3$ with $8$ processes, $128^3$ with $64$ processes, and $256^3$ with $512$ processes. Both axes are log scales. %
  }
  
  \label{fig:weak_scaling}
\end{figure}

\subsubsection{Weak Scalability Study}

We measure the influence of using asynchronous parallelism on weak scaling of distributed union-find. The weak scaling evaluates the performance when each process is assigned a constant-size problem as the number of processes increases. In this study, each process is assigned with a $32^3$ mesh grid with a constant feature density. 
Test data with sizes of $32^3$, $64^3$, $128^3$, and $256^3$ are processed using $1$, $8$, $64$, and $512$ processes respectively. 
The ideal weak scaling is when the execution time is constant in all the runs.

Compared with the baseline using bulk-synchronous parallelism, the asynchronous parallelism leads to significant enhancement in weak scaling efficiency. 
The results are displayed in Fig.~\ref{fig:weak_scaling}. 
When using $512$ processes, the baseline achieves weak scaling efficiency of $0.74\%$ in the critical point tracking benchmark and $1.39\%$ in the super-level set extraction benchmark. Our distributed union-find with the asynchronous parallelism attains $8.13\%$ and $20.66\%$ efficiency in the two benchmarks, respectively, with a speedup of $10.59$x and $13.66$x over the baseline.

\subsection{Scientific Applications} \label{sect:applications}

We evaluate distributed union-find, especially the load balancing performance, in two scientific applications: (1) tracking critical points in exploding wire experimental data and (2) extracting super-level sets in fusion plasma simulation data. Both of the scientific data are time-varying. 

\subsubsection{Application Background and Benchmark Setting}

We give background and benchmark settings for the two scientific applications.

\textbf{Exploding wire experiments: }
Scientists can use an exploding-wire apparatus to generate many high-temperature microparticles \cite{wang2016four, wang2020microparticle}. 
High-speed imaging cameras can capture the movement of these particles and produce high-resolution images. Tracking these particles on the images helps physicists understand the particles' physical properties and helps computational scientists enhance theoretical models for simulation development. 



\begin{figure}[htb]
\centering
\captionsetup[subfigure]{labelformat=empty}

  \begin{subfigure}[b]{0.08\linewidth}
   \includegraphics[width=\linewidth]{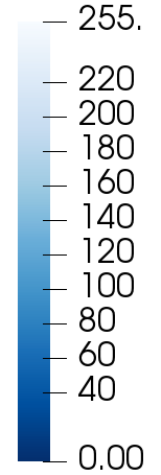}
   \caption{}
 \end{subfigure}
 \begin{subfigure}[b]{0.44\linewidth}
   \includegraphics[width=\linewidth]{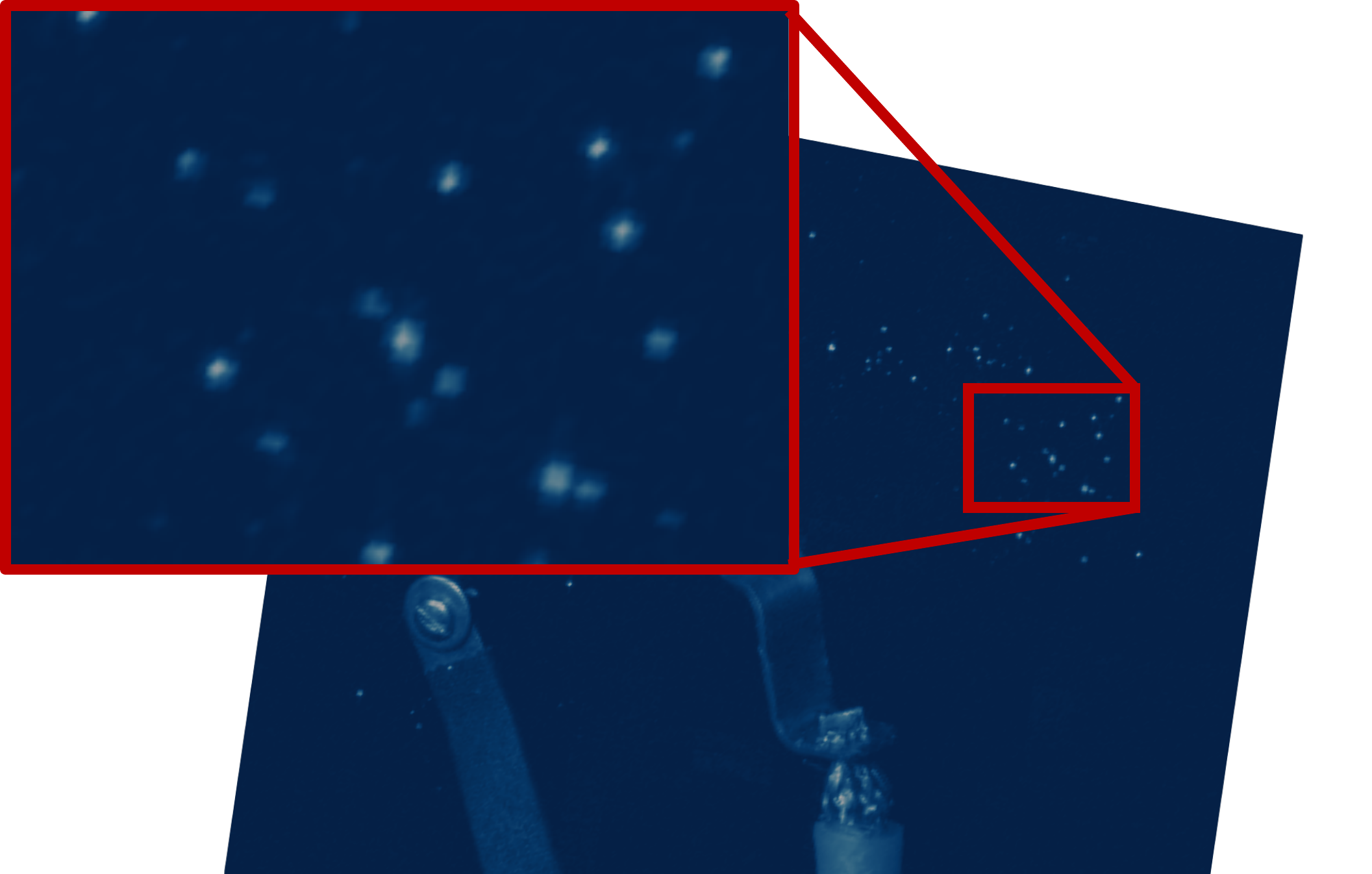}
   \caption{(a)}
 \end{subfigure}
 \begin{subfigure}[b]{0.44\linewidth}
   \includegraphics[width=\linewidth]{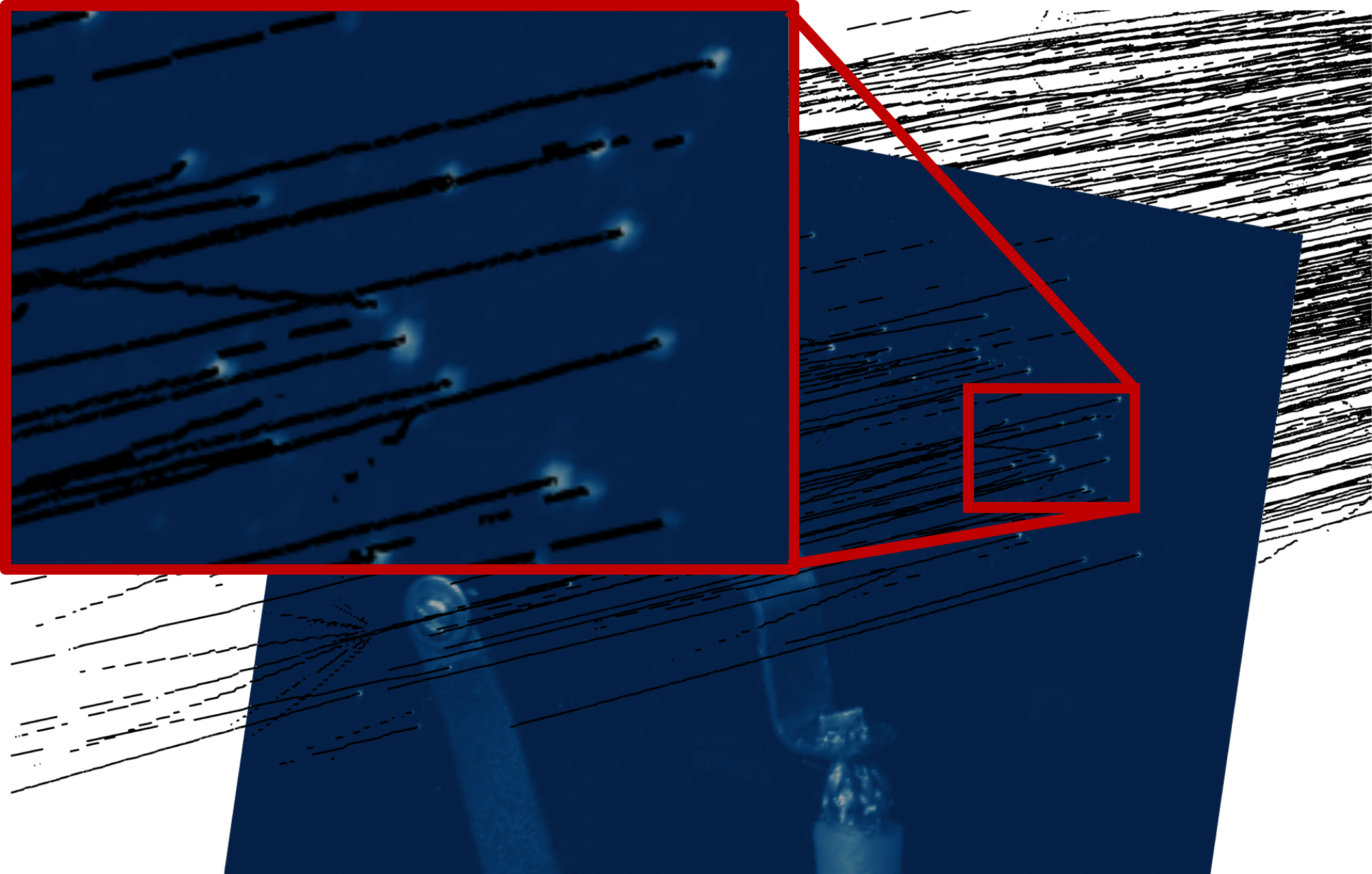}
   \caption{(b)}
 \end{subfigure} 

  \caption{Tracking critical points in the exploding wire experimental data. The intensity value ranges from 0 to 255. 
  (a): One image frame, where bright particles are detected as maximum points. 
  (b): Trajectories of maximum points represented by black lines.  
  }
  
  \label{fig:exploding_wire_illustration}
\end{figure}

We model particles as local maximum points in each frame and track the points' movement across frames of the exploding wire imaging data. A frame of the time-varying data is shown in Fig.~\ref{fig:exploding_wire_illustration}a. 
The test data have a $384\times384$ spatial resolution and $4,745$ timesteps. $3,197,333$ maximum points are detected from the data. The maximum points' trajectories are shown in Fig.~\ref{fig:exploding_wire_illustration}b, which pass through the particles on the image. 



\begin{figure}[htb]
\centering
\captionsetup[subfigure]{labelformat=empty}

 \begin{subfigure}[b]{0.08\linewidth}
   \includegraphics[width=\linewidth]{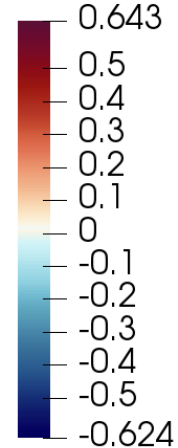}
   \caption{}
 \end{subfigure}
 \begin{subfigure}[b]{0.39\linewidth}
   \includegraphics[width=\linewidth]{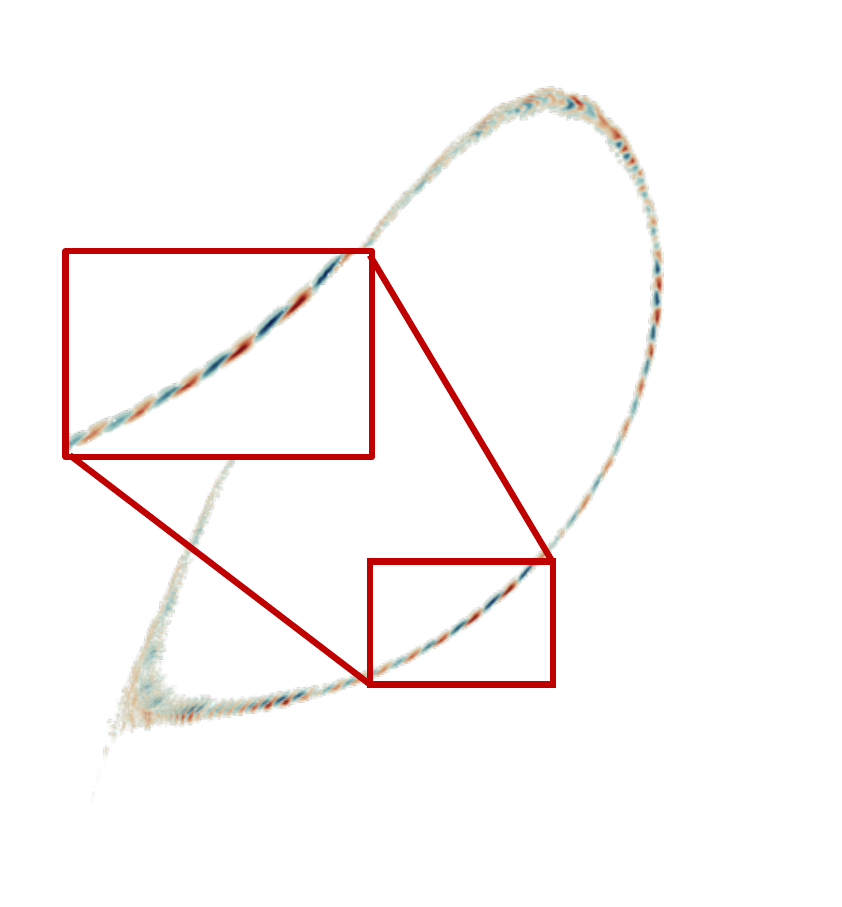}
   \caption{(a)}
 \end{subfigure}
 \begin{subfigure}[b]{0.39\linewidth}
   \includegraphics[width=\linewidth]{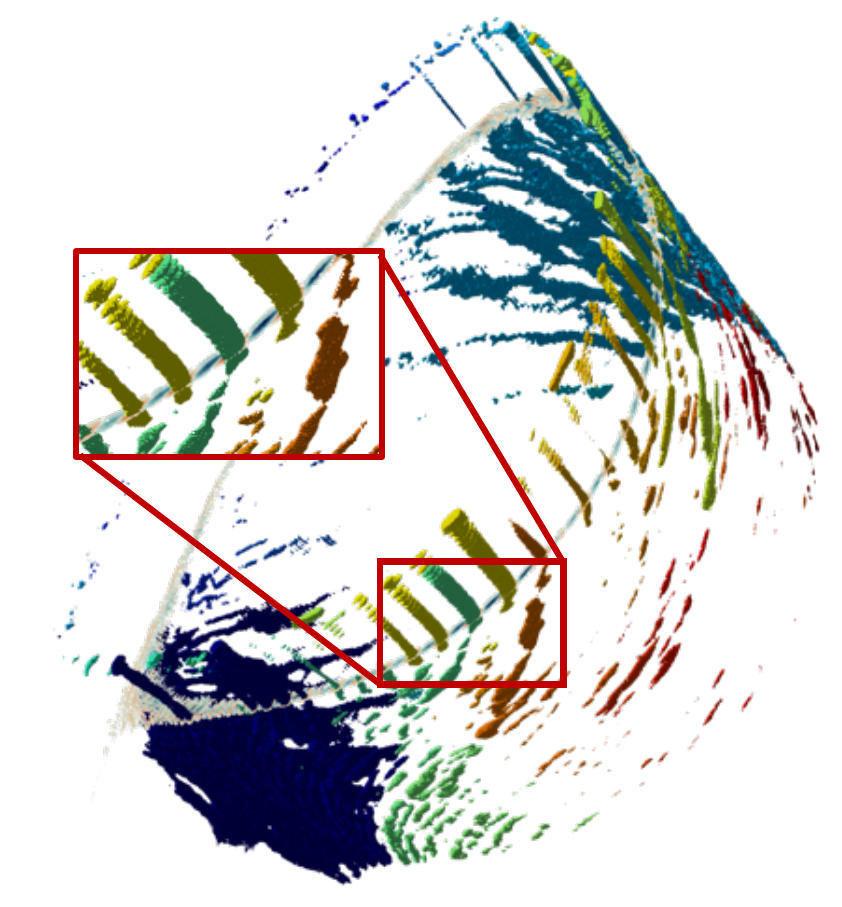}
   \caption{(b)}
 \end{subfigure} 
 \begin{subfigure}[b]{0.08\linewidth}
   \includegraphics[width=\linewidth]{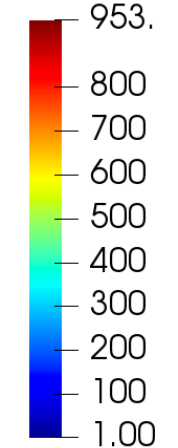}
   \caption{}
 \end{subfigure}

  \caption{Tracking super-level sets in fusion plasma simulation data. 
  (a): 2D density field at a timestep, where blobs are high-density regions and are detected as super-level sets. 
  (b): Extracted super-level sets having $953$ labeled connected components colored by unique hues. 
  }
  
  \label{fig:bout_illustration}
\end{figure}

\textbf{Fusion plasma simulations: }
A Tokamak torus device is the mainstream fusion reactor to confine plasma magnetically in order to achieve fusion energy production magnetically. The turbulent transport from the edge plasma usually takes a ubiquitous form of filaments, defined as density-enhancement coherent structures, also referred to as blobs. 
The blob movement may cause loss of plasma and severe damage to the device, and hence, has been subject to intensive researches~\cite{KRASHENINNIKOV2001,Russell, nespoli20193d}. 



We track blobs in ion density fluctuation data produced by a BOUT++ electromagnetic fluid simulation \cite{xqxu2000pop,DUDSON2009} for the fusion reactor. 
A separatrix slice of the data is shown in Fig.~\ref{fig:bout_illustration}a. 
Blobs usually are the regions of high ion density. Hence, following the work of \cite{nespoli20193d}, we model blobs as regions with densities larger than $2.5$ standard deviation than the average density and track super-level sets across timesteps. 
The test data have $425\times880$ spatial resolution and $701$ timesteps. $1,708,341$ high-density voxels are extracted from the data. The results are shown in Fig.~\ref{fig:bout_illustration}b. 



\begin{figure}[htb]
\centering
\includegraphics[width=\linewidth]{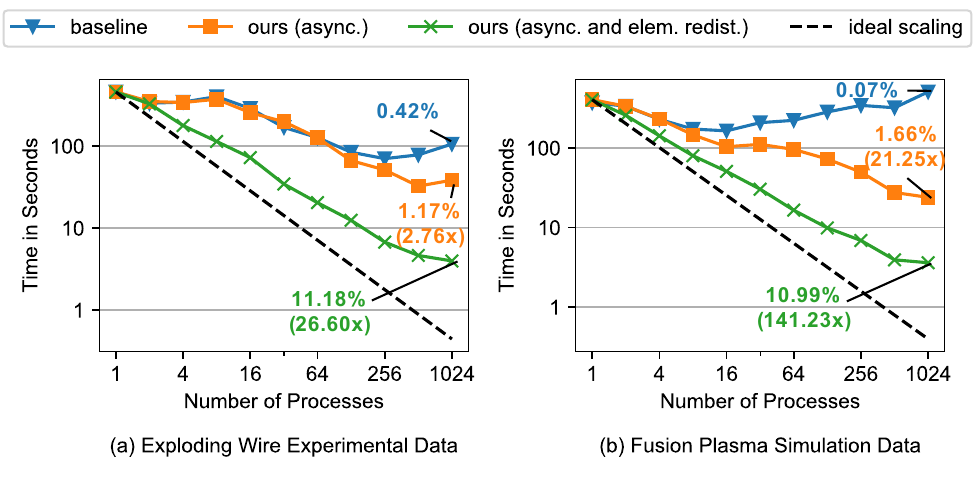}

  \caption{Strong scaling of distributed union-find for tracking and extracting features in two application datasets: (a) exploding wire experimental data and (b) fusion plasma simulation data. Both axes are log scales. 
  We compare a baseline (distributed union-find of Iverson et al. \cite{iverson2015evaluation} with balanced mesh cells \cite{harrison2015}) with our distributed asynchronous union-find without/with the redistribution of feature elements. 
  }
  
  \label{fig:scientific_data_union_find}
\end{figure}

\subsubsection{Benchmark in Scientific Applications}


We demonstrate the performance of distributed union-find in the two scientific applications. 
Fig.~\ref{fig:scientific_data_union_find} displays the strong scaling results. 
When $1,024$ processes are used, the bulk-synchronous baseline approach achieves strong scaling efficiency of $0.42\%$ in the exploding wire data and $0.07\%$ in the fusion plasma data. The asynchronous parallelism improves the efficiency to $1.17\%$ and $1.66\%$ with a speedup of $2.76$x and $21.25$x over the baseline, respectively. However, the scalability is still limited by the imbalanced features. 

\begin{figure}[htb]
\centering
\includegraphics[width=\linewidth]{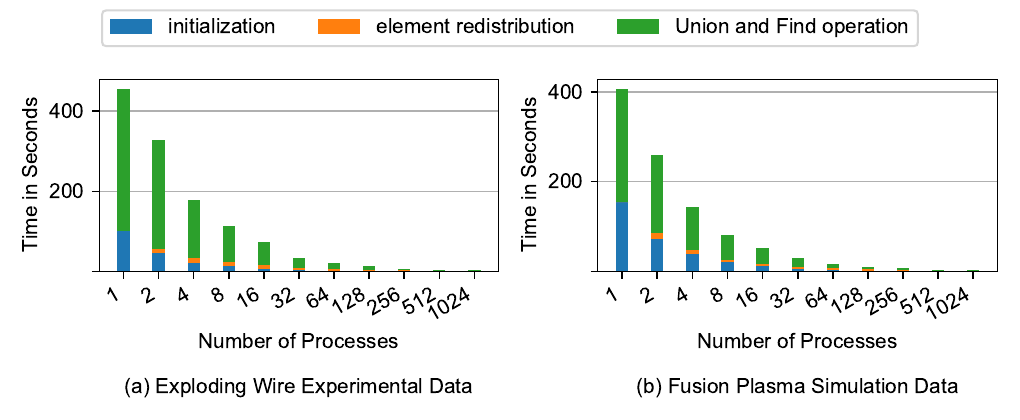}

  \caption{
  Breakdown of the time cost of our distributed union-find algorithm with both the asynchronous parallelism and the element redistribution in two application datasets. The horizontal axis is a log scale. The ``initialization'' includes the initial assignment of element IDs and the initialization of data structures. 
  }
  
  \label{fig:scientific_data_union_find_breakdown}
\end{figure}

\textbf{Evaluation of load balancing: }
We evaluate the load balancing of distributed union-find using the k-d tree based element redistribution in the two scientific applications. 
As visualized in Fig.~\ref{fig:exploding_wire_illustration} and Fig.~\ref{fig:bout_illustration}, the features are not uniformly distributed in domain for both of the scientific applications. 
Hence, balancing mesh cells may not be effective enough for these cases, and balancing feature elements is expected to improve the distributed union-find performance. 
After redistributing feature elements to balance the number of features in each process, the strong scaling efficiency increases significantly to $11.18\%$ and $10.99\%$, respectively. Compared with the baseline, our distributed asynchronous union-find with the feature element redistribution attains $26.60$x speedup for tracking critical points in the exploding wire data and $141.23$x speedup for tracking super-level sets in the fusion plasma data. 
Fig.~\ref{fig:scientific_data_union_find_breakdown} displays the cost breakdown of our distributed asynchronous union-find with the feature element redistribution.

\section{Discussions} 
\label{sect:discussion}

We discuss limitations of our distributed union-find. 


\textbf{Limitation of element identifier: }
Our distributed asynchronous union-find assumes elements have unique and sortable IDs, which may not be the case for all datasets. In scientific applications using mesh data, we use IDs of mesh cells or mesh faces as elements' IDs, where the cost of such ID assignment is included in ``initialization'' of Fig.~\ref{fig:scientific_data_union_find_breakdown}.  However, for a graph structure without element IDs as input, an additional preprocessing for element ID assignment is required. A possible way to remedy this could be to collect element counts of all processes, compute a numeric ID range for each process, and each process assigns IDs to local elements in parallel within the ID range, which introduces additional cost.


\textbf{Limitation of memory capacity: }
Out-of-core algorithms may be needed if the memory capacity cannot hold a single data block. 
A possible solution to remedy the limitation is described in the following. 
First, we decompose the data further into smaller data blocks with ghost layers such that the memory of each process is able to hold a single data block. Each process then loads a smaller data block. 
Second, each process extracts elements and edges of interest in the loaded data block to perform distributed union-find. 
Third, processes release current blocks and load new unprocessed blocks. We repeat the second and the third step until all data blocks are processed.

\section{Conclusion and Future Work}
\label{sect:conclusion}

This paper presents a novel distributed union-find algorithm that (1) overlaps communications with computations using asynchronous parallelism to reduce synchronization costs and (2) redistributes set elements using a distributed k-d tree decomposition to balance processes' workloads for scalable scientific visualization and analysis. 
Our algorithm demonstrated improved scaling characteristics than existing distributed union-find methods in the scientific applications of critical point tracking and super-level set extraction. 
Benchmark datasets included synthetic data, exploding wire experimental data, and fusion plasma simulation data. 

In the future, first, we will evaluate our algorithm's performance in 4D (3D in space and 1D in time) scientific data. 
Distributed union-find algorithms can be extended to 4D seamlessly because the union-find algorithms accept elements and edges between elements as input, independent of the dimensionality of elements. Also, the element redistribution using distributed k-d trees can deal with elements with higher dimensions. 
Second, we will integrate our algorithm into other visualization and analysis applications, such as in situ visualization and graph/network data analysis.


%




\ifCLASSOPTIONcompsoc
  \section*{Acknowledgments}
\else
  \section*{Acknowledgment}
\fi

The authors would like to thank Eric Brugger from Lawrence Livermore National Laboratory for providing the help of preprocessing fusion plasm simulation data. 
This work is supported in part by National Science Foundation Division of Information and Intelligent Systems-1955764, U.S. Department of Energy Los Alamos National Laboratory contract 47145, and UT-Battelle LLC contract 4000159447 program manager Laura Biven. 
The research is also supported by the Exascale Computing Project (ECP), project number 17-SC-20-SC, a collaborative effort of Department of Energy Office of Science and the National Nuclear Security Administration, as part of the Co-design center for Online Data Analysis and Reduction (CODAR)~\cite{CODAR2020}. 
It is also supported by the U.S. Department of Energy, Office of Advanced Scientific Computing Research, Scientific Discovery through Advanced Computing (SciDAC) program, and by Laboratory Directed Research and Development (LDRD) funding from Argonne National Laboratory, provided by the Director, Office of Science, of the U.S. Department of Energy under Contract No. DE-AC02-06CH11357.  This research used resources of the Argonne Leadership Computing Facility, which is a DOE Office of Science User Facility supported under Contract DE-AC02-06CH11357.

\ifCLASSOPTIONcaptionsoff
  \newpage
\fi



\bibliographystyle{IEEEtran}
\bibliography{union_find}
%



%




\begin{IEEEbiography}[{\includegraphics[width=1in,height=1.25in,clip,keepaspectratio]{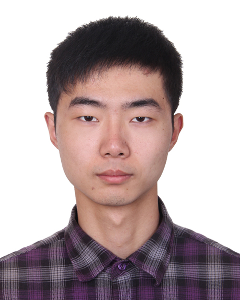}}]{Jiayi Xu} 
is a Ph.D. candidate in the Department of Computer Science and Engineering at the Ohio State University. His research interests include graph visualization and scientific feature tracking. Xu received his B.S. degree in computer science and technology from Chu Kochen Honors College at Zhejiang University in 2014. 
\end{IEEEbiography}


\begin{IEEEbiography}[{\includegraphics[width=1in,height=1.25in,clip,keepaspectratio]{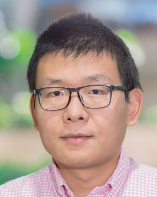}}]{Hanqi Guo} 
is an assistant computer scientist at Argonne National Laboratory, scientist at the University of Chicago Consortium for Advanced Science and Engineering (CASE), and fellow of the Northwestern Argonne Institute for Science and Engineering (NAISE).  His research interests include data analysis, visualization, and machine learning for scientific data.  He has published more than 40 research papers in top visualization journals and conferences including IEEE VIS, IEEE TVCG, and IEEE TPDS.  He is also the recipient of the best paper award in IEEE VIS 2019 and the winner of the 2017 Postdoctoral Performance Award in Basic Research in Argonne National Laboratory.  He received his Ph.D. degree in computer science from Peking University in 2014 and his B.S. degree in mathematics and applied mathematics from Beijing University of Posts and Telecommunications in 2009. 
\end{IEEEbiography}


\begin{IEEEbiography}[{\includegraphics[width=1in,height=1.25in,clip,keepaspectratio]{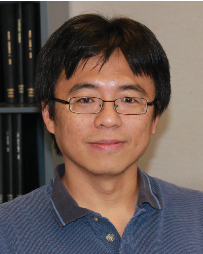}}]{Han-Wei Shen} 
is a full professor at the Ohio State University. He received his B.S. degree from the Department of Computer Science and Information Engineering at National Taiwan University in 1988, his M.S. degree in computer science from the State University of New York at Stony Brook in 1992, and his Ph.D. degree in computer science from the University of Utah in 1998. From 1996 to 1999, he was a research scientist at NASA Ames Research Center in Mountain View California. His primary research interests are scientific visualization and computer graphics.  He is a winner of the National Science Foundation's CAREER award and U.S. Department of Energy's Early Career Principal Investigator Award. He also won the Outstanding Teaching award twice in the Department of Computer Science and Engineering at the Ohio State University.
\end{IEEEbiography}


\begin{IEEEbiography}[{\includegraphics[width=1in,height=1.25in,clip,keepaspectratio]{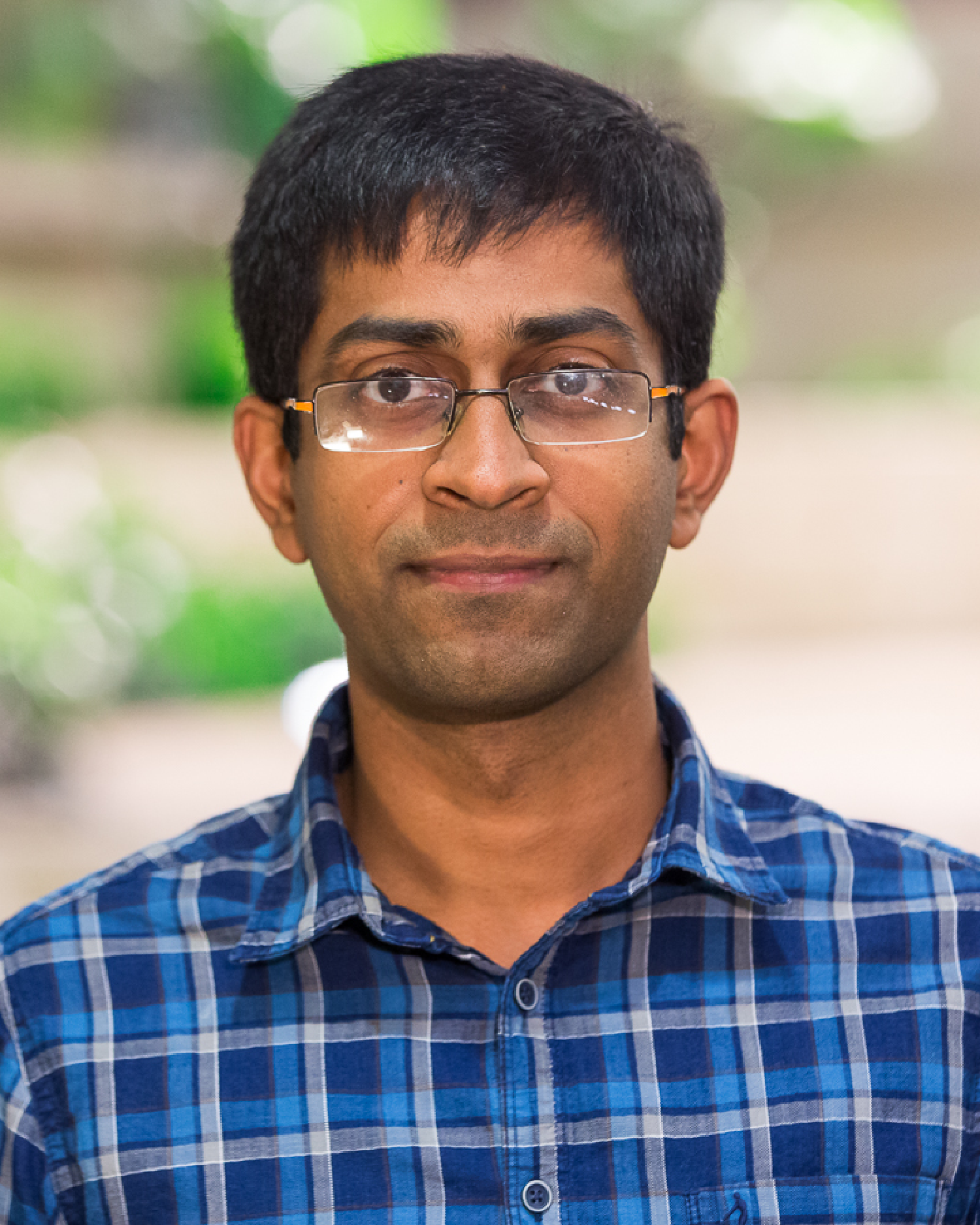}}]{Mukund Raj} 
is a postdoctoral appointee in the Mathematics and Computer Science Division, Argonne National Laboratory. His research interests include ensemble visualization, scientific visualization, and in situ analysis. Raj received his PhD in computing from the University of Utah in 2018. 
\end{IEEEbiography}


\begin{IEEEbiography}[{\includegraphics[width=1in,height=1.25in,clip,keepaspectratio]{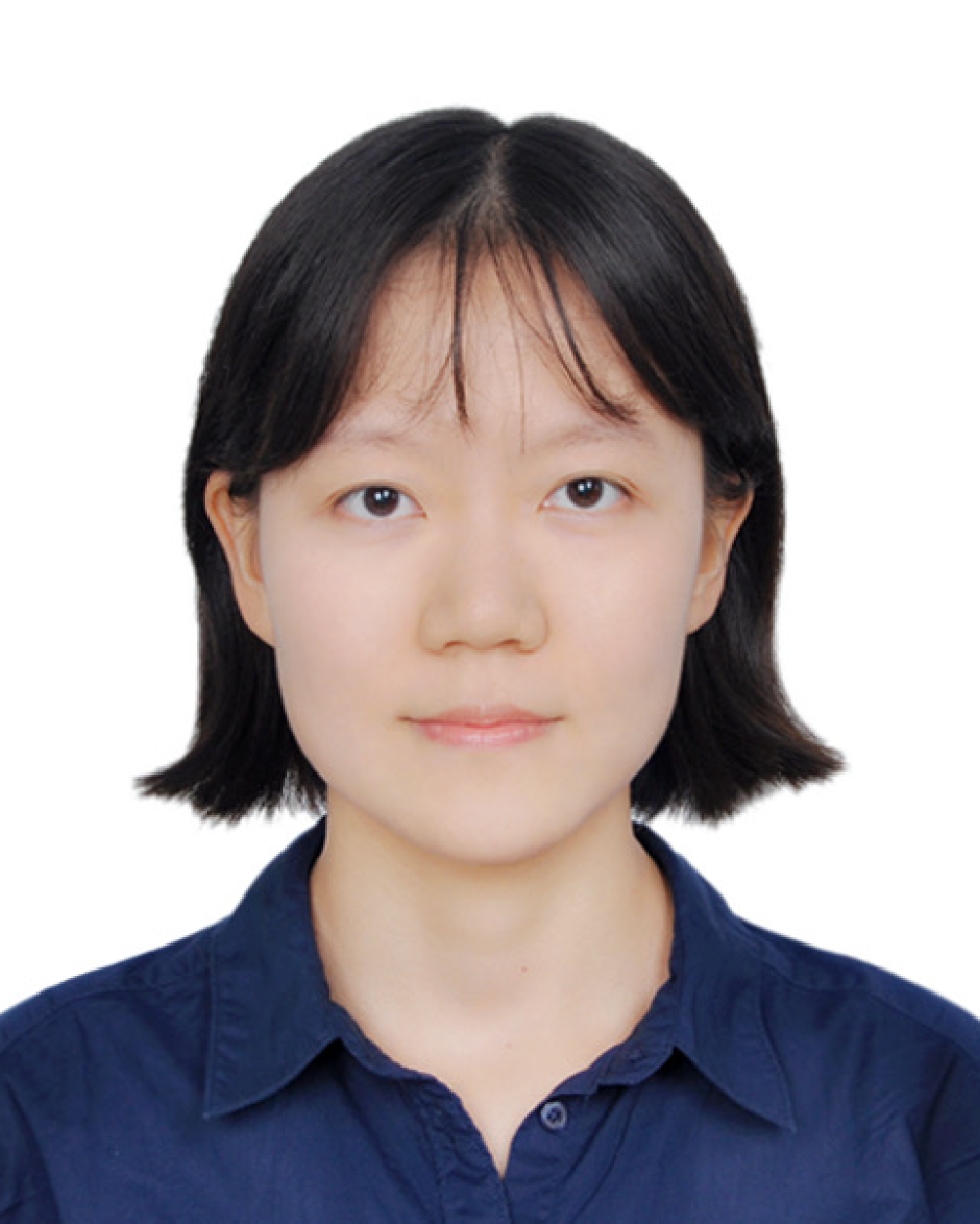}}]{Xueyun Wang} 
Xueyun Wang received her B.S. degree in 2015 and Ph.D. degree in 2020 from School of Physics, Peking University, Beijing, China. Her major interests include fluid simulation of plasma physics.
\end{IEEEbiography}


\begin{IEEEbiography}[{\includegraphics[width=1in,height=1.25in,clip,keepaspectratio]{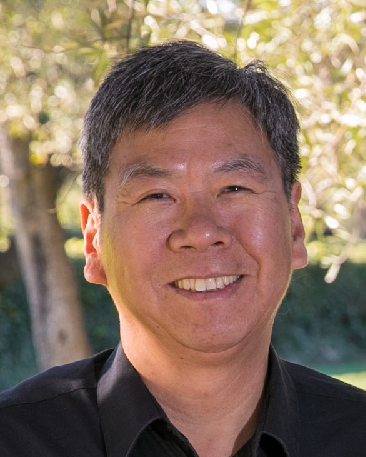}}]{Xueqiao Xu} 
is a principal physicist at Lawrence Livermore National Laboratory. His research interests include plasma physics and controlled nuclear fusion, plasma theory and large scale simulations, and machine learning for moment closures of kinetic equations. He has published more than 160 research papers in top plasma physics journals that include PRL, Nuclear Fusion, POP, PPCF, JCP, CPC, CiCP. Xu received his Ph.D. in physics from the University of Texas at Austin in 1990.
\end{IEEEbiography}


\begin{IEEEbiography}[{\includegraphics[width=1in,height=1.25in,clip,keepaspectratio]{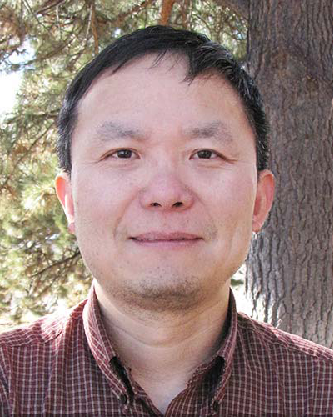}}]{Zhehui Wang} 
a physicist at Los Alamos National Laboratory (LANL). He received his B.S. degree from the Department of the Earth and Space Sciences at the University of Science and Technology of China in 1992, his M.S.  and Ph. D degrees in Astrophysical Sciences (Plasma Physics) from Princeton in 1994 and1998 respectively. He joined LANL as a postdoc in 1998, and has been a staff member since 2001. One of his recent research interests is applied data science to problems in imaging of physical systems and particle tracking. He has authored and coauthored more than 100 peer-reviewed papers and holds six US patents in plasma accelerator, velocimetry, neutron detectors and X-ray cameras. He currently leads the high-speed imaging team funded by several DOE and internal LANL programs.
\end{IEEEbiography}


\begin{IEEEbiography}[{\includegraphics[width=1in,height=1.25in,clip,keepaspectratio]{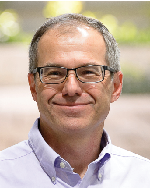}}]{Tom Peterka} 
is a computer scientist at Argonne National Laboratory, a scientist at the University of Chicago Consortium for Advanced Science and Engineering (CASE), an adjunct assistant professor at the University of Illinois at Chicago, and a fellow of the Northwestern Argonne Institute for Science and Engineering (NAISE). His research interests are in large-scale parallel in situ analysis of scientific data. Recipient of the 2017 DOE Early Career Award and four best paper awards, Peterka has published over 100 peer-reviewed papers in conferences and journals that include ACM/IEEE SC, IEEE IPDPS, IEEE VIS, IEEE TVCG, and ACM SIGGRAPH. Peterka received his Ph.D. in computer science from the University of Illinois at Chicago in 2007, and he currently leads several DOE- and NSF-funded projects.
\end{IEEEbiography}








\end{document}